\documentclass[aps,showpacs,preprintnumbers,amsmath, amssymb]{revtex4}

\usepackage{float}
\usepackage{graphics,epsfig}
\usepackage{graphicx}
\usepackage{dcolumn}
\usepackage{bm}

\begin{document}
\baselineskip=0.8 cm
\title{{\bf Generalized superconductors from the coupling of a scalar field to the Einstein tensor and their refractive
index in massive gravity}}

\author{Manman Sun$^{1}$, Dong Wang$^{1}$, Qiyuan Pan$^{1,2}$\footnote{panqiyuan@126.com}, and Jiliang Jing$^{1,2}$\footnote{jljing@hunnu.edu.cn}}
\affiliation{$^{1}$Key Laboratory of Low Dimensional Quantum
Structures and Quantum Control of Ministry of Education, Synergetic
Innovation Center for Quantum Effects and Applications, and
Department of Physics, Hunan Normal University, Changsha, Hunan
410081, China} \affiliation{$^{2}$Center for Gravitation and
Cosmology, College of Physical Science and Technology, Yangzhou
University, Yangzhou 225009, China}

\vspace*{0.2cm}
\begin{abstract}
\baselineskip=0.6 cm
\begin{center}
{\bf Abstract}
\end{center}

We construct the generalized superconductors from the coupling of a
scalar field to the Einstein tensor in the massive gravity and
investigate their negative refraction in the probe limit. We observe
that the larger graviton mass and Einstein tensor coupling
parameters both hinder the formation of the condensation, but the
larger graviton mass or smaller coupling parameter makes it easier
for the emergence of the Cave of Winds. Furthermore, we see that the
larger graviton mass but smaller coupling parameter make the range
of frequencies or the range of temperatures larger for which a
negative Depine-Lakhtakia index occurs, which indicates that the
graviton mass and Einstein tensor have completely different effects
on the negative refraction. In addition, we find that the larger
graviton mass and coupling parameters both can reduce the
dissipation and improve the propagation in the holographic setup.

\end{abstract}
\maketitle
\newpage
\vspace*{0.2cm}

\section{Introduction}

It is well known that Maldacena conjectured a remarkable connection
between a strongly coupled field theories to a gravitational theory
in higher spacetime dimensions in 1998 \cite{Maldacena}, which
states that a $d$-dimensional weakly coupled dual gravitational
description in the bulk is equivalent to a $(d-1)$-dimensional
strongly coupled conformal field theory on the boundary. This is the
so-called anti-de Sitter/conformal field theories (AdS/CFT)
correspondence \cite{Witten,Gubser1998}, which provides a novel
approach to studying the strongly coupled systems in condensed
matter physics in the past ten years, especially the high
temperature superconductors \cite{HartnollPRL101}. With this
holographic duality, it is shown that the holographic superconductor
models exhibit many characteristic properties shared by real
superconductors \cite{JZaanen}; for reviews, see Refs.
\cite{HartnollRev,HerzogRev,HorowitzRev,CaiRev} and references
therein. However, in most cases, the studies on the holographic
superconductors focus on the superconductors without an impurity.
Turning on a coupling between the gauge field and a new massive
gauge field, the authors of Ref. \cite{Ishii} analyzed the impurity
effect in a holographic superconductor and observed that the mass
gap in the optical conductivity disappears when the coupling is
sufficiently large. More recently, Hu \emph{et al.} constructed the
holographic Josephson junction from the massive gravity and found
that the increasing graviton mass parameter will change the
superconductor phase into a normal metal phase for a fixed
temperature (chemical potential), which indicates that the graviton
mass parameter and the doping have a similar effect as from
superconductivity to a normal metal \cite{HuPRD2016}. Considering a
holographic superconductor with a scalar field coupled kinematically
to the Einstein tensor, Kuang and Papantonopoulos observed that, as
the strength of the coupling increases, the critical temperature
below which the scalar field condenses is lowering, the condensation
gap decreases faster than the temperature, the width of the
condensation gap is not proportional to the size of the condensate
and at low temperatures the condensation gap tends to zero for the
strong coupling, which suggests that the derivative coupling in the
gravity bulk can have a dual interpretation on the boundary
corresponding to impurities concentrations in a real material
\cite{KuangE2016}. Other investigations based on the impurity effect
on the holographic dual models can be found, for example, in Refs.
\cite{ZengZhang,FangJHEP}.

On the other hand, since Veselago proposed in theory that the
refractive index might be negative in some special material
\cite{Veselago}, there have been many investigations concerning this
exotic electromagnetic phenomenon---negative refraction, which
indicates that electromagnetic waves propagate in a direction
opposite to that of the flow of energy in the so-called
``metamaterials" \cite{Smith,Pendry,Ramakrishna}. Interestingly, it
is believed that the superconductors are the ideal candidates for
metamaterials due to their low loss, compact structure,
extraordinary degree of nonlinearity and tunability, magnetic flux
quantization and the Josephson effect, quantum effects in which
photons interact with quantized energy levels in the meta-atom, and
strong diamagnetism \cite{Anlage,JungUA}. With the help of the
AdS/CFT correspondence, the refractive index in the superconductors
was investigated by holography in the probe limit \cite{GaoZhang}
and away from the probe limit \cite{AmaritiFMS2011}, which shows
that in the superconducting phase there is negative refraction at
low enough frequencies only in the backreacted case. It is
interesting to note that Mahapatra \emph{et al.}
\cite{MahapatraJHEP2014} constructed the generalized holographic
superconductors via the St\"{u}ckelberg mechanism
\cite{Franco,FrancoJHEP,PanWangPLB2010} in the four dimensional
R-charged black hole and observed that the superconducting phase can
exhibit a negative Depine-Lakhtakia (DL) index
\cite{DepineLakhtakia} at low frequencies and below a cut-off value
of the charge parameter even in the probe limit. Extending the
investigation to the generalized superconductors with Born-Infeld
electrodynamics, Cheng \emph{et al.} found that the system has a
negative DL index in the superconducting phase at small frequencies
and the greater the Born-Infeld corrections the larger the range of
frequencies or the range of temperatures for which the negative
refraction occurs \cite{ChengEPJC2018}. The refractive index in the
holographic dual models can also be found, for example, in Refs.
\cite{AmaritiFMP2011,GJSJHEP2011,LiuPRD2011,AmaritiFMP2013,JiangPRD2013,
PhukonSJHEP2013,DeyMT2014,ForcellaJHEP2014,MahapatraJHEP2015,JiangPRD2016,Zangeneh}.

In this work, motivated by the recent studies in Refs.
\cite{HuPRD2016,KuangE2016}, we will use the AdS/CFT correspondence
to investigate systematically the effect of the doping or impurity
on the superconductors in strongly coupled condensed matter systems
and their negative refraction in the probe limit. In order to obtain
the rich phase structure, we construct the generalized
superconductors from the coupling of a scalar field to the Einstein
tensor in the massive gravity via the St\"{u}ckelberg mechanism just
as in \cite{MahapatraJHEP2014}. We will observe that the larger
graviton mass and Einstein tensor coupling parameters both hinder
the formation of the condensation, but the larger graviton mass or
smaller coupling parameter makes it easier for the emergence of the
so-called Cave of Winds \cite{Arean2010}, i.e., the second-order
transition occurs before the first-order transition to a new
superconducting phase when the temperature decreases
\cite{PWavSuperfluid,LifshitzSuperfluid,GBSuperfluid}. Furthermore,
we assume that the boundary theory is weakly coupled to a dynamical
electromagnetic field and one can calculate the refractive index of
the system perturbatively. We will find that the larger graviton
mass parameter or smaller coupling parameter makes the range of
frequencies or the range of temperatures larger for which a negative
DL index occurs, but the larger graviton mass and coupling
parameters both can reduce the dissipation and improve the
propagation in the holographic setup.

The organization of this work is as follows. In Sec. II we will use
the AdS/CFT correspondence to construct the generalized
superconductors from the coupling of a scalar field to the Einstein
tensor in the massive gravity. In Sec. III we will discuss the
effects of the graviton mass and Einstein tensor coupling parameters
on the scalar condensation and phase transition in the generalized
holographic models. In Sec. IV we will analyze the effects of the
graviton mass and Einstein tensor coupling parameters on the
negative refraction in the generalized holographic models. We will
conclude in the last section of our main results.

\section{Generalized superconductors from the coupling of a scalar field to the Einstein tensor in massive gravity}

Since we are interested in the background of a massive gravity
theory, we start with the following action for a four-dimensional
ghost-free dRGT massive gravity \cite{Vegh,CaiHuPZ}
\begin{equation}
\label{action} S =\frac{1}{16\pi G}\int d^{4}x \sqrt{-g} [R
+\frac{6}{L^2}+m^2 \sum^4_i c_i {\cal U}_i (g,\textrm{f})],
\end{equation}
where $\textrm{f}$ is the reference metric, $m$ is the graviton mass
parameter, $c_i$ are constants and ${\cal U}_i$ are symmetric
polynomials of the eigenvalues of the matrix ${\cal K}^{\mu}_{\ \nu}
\equiv \sqrt {g^{\mu\alpha}\textrm{f}_{\alpha\nu}}$
\begin{eqnarray}
\label{action-B}
&& {\cal U}_1= [{\cal K}], \nonumber \\
&& {\cal U}_2= [{\cal K}]^2 -[{\cal K}^2], \nonumber \\
&& {\cal U}_3= [{\cal K}]^3 - 3[{\cal K}][{\cal K}^2]+ 2[{\cal K}^3], \nonumber \\
&& {\cal U}_4= [{\cal K}]^4- 6[{\cal K}^2][{\cal K}]^2 + 8[{\cal
K}^3][{\cal K}]+3[{\cal K}^2]^2 -6[{\cal K}^4],
\end{eqnarray}
where the square root in ${\cal K}$ means $(\sqrt{A})^{\mu}_{\
\nu}(\sqrt{A})^{\nu}_{\ \lambda}=A^{\mu}_{\ \lambda}$ and $[{\cal
K}]=K^{\mu}_{\ \mu}$. The equation of motion of this action will be
\begin{eqnarray}
R_{\mu\nu}-\frac{1}{2}Rg_{\mu\nu}-\frac{3}{L^2} g_{\mu\nu}+m^2
\chi_{\mu\nu}=8\pi GT_{\mu\nu},
\end{eqnarray}
with
\begin{eqnarray}
\chi_{\mu\nu}&=&-\frac{c_1}{2}({\cal U}_1g_{\mu\nu}-{\cal
K}_{\mu\nu})-\frac{c_2}{2}({\cal U}_2g_{\mu\nu}-2{\cal U}_1{\cal
K}_{\mu\nu}+2{\cal K}^2_{\mu\nu}) -\frac{c_3}{2}({\cal
U}_3g_{\mu\nu}-3{\cal U}_2{\cal K}_{\mu\nu}+6{\cal U}_1{\cal
K}^2_{\mu\nu}-6{\cal K}^3_{\mu\nu}) \nonumber \\
&&-\frac{c_4}{2}({\cal U}_4g_{\mu\nu}-4{\cal U}_3{\cal
K}_{\mu\nu}+12{\cal U}_2{\cal K}^2_{\mu\nu}-24{\cal U}_1{\cal
K}^3_{\mu\nu}+24{\cal K}^4_{\mu\nu}).
\end{eqnarray}
We consider the following reference metric \cite{Vegh}
\begin{equation}
\label{reference} \textrm{f}_{\mu\nu} = {\rm diag}(0,0, c_0^2
h_{ij}),
\end{equation}
with $c_0$ being a positive constant, and a general black hole
solution is given by \cite{CaiHuPZ}
\begin{equation}\label{MassiveBH}
ds^2=-r^2f\left(r\right)dt^2+\frac{1}{r^2f\left(r\right)}dr^2+r^2h_{ij}dx^{i}dx^{j},
\end{equation}
with
\begin{eqnarray}
f(r)=\frac{\kappa}{r^{2}}+\frac{1}{L^{2}}-\frac{m_{0}}{r^{3}}+\frac{c_{1}m^{2}}{2r}+\frac{c_{2}m^{2}}{r^{2}},
\end{eqnarray}
where $m_{0}$ is related to the mass of the black hole, and
$h_{ij}dx^{i}dx^{j}$ is the line element for the two-dimensional
spherical, flat or hyperbolic space with $\kappa=-1,~0$ or $1$,
respectively. In this work we are only interested in the
plane-symmetric black hole, so we will set $\kappa=0$ and the
corresponding Hawking temperature is determined by
\begin{eqnarray}
T_{H}=\frac{1}{4\pi
r_{h}}\left(\frac{3r^{2}_{h}}{L^{2}}+c_{1}m^{2}r_{h}+c_{2}m^{2}\right),
\end{eqnarray}
where $r_{h}$ is the radius of the event horizon. For this
considered solution, the nonzero components of the Einstein tensor
$G^{\mu\nu}$ are
\begin{eqnarray}\label{EinsteinTensor}
G^{tt}=-\frac{1}{r^{2}}\left(3+\frac{rf^{\prime}}{f}\right),~~
G^{rr}=r^{2}f^{2}\left(3+\frac{rf^{\prime}}{f}\right),~~G^{xx}=G^{yy}=\frac{1}{2r^{2}}\left[6f+r(6f^\prime+rf'')\right],
\end{eqnarray}
where the prime denotes a derivative with respect to $r$. Without
loss of generality, we will scale the AdS radius $L=1$ in the
following calculation.

In the probe limit, we consider a Maxwell field and a charged
complex scalar field coupled to the Einstein tensor $G^{\mu\nu}$
with the generalized action
\begin{equation}
S=\int d^{4}x\sqrt{-g}\left[-\frac{F_{\mu\nu}F^{\mu\nu}}{4}-
\frac{(g^{\mu\nu}+\eta
G^{\mu\nu})(\partial_{\mu}\Psi)(\partial_{\nu}\Psi)}{2}
-\frac{m_{\Psi}^2\Psi^2}{2}-|\textrm{G}(\Psi)|(g^{\mu\nu}+\eta
G^{\mu\nu}) (\partial_{\mu}\alpha-A)(\partial_{\nu}\alpha-A)\right],
\label{Lagrangian}
\end{equation}
where the charged scalar field $\Psi$ with the mass $m_{\Psi}$ and
the phase $\alpha$ both are real, and the local $U(1)$ gauge
symmetry is given by $A_{\mu}\rightarrow
A_{\mu}+\partial_{\mu}\lambda$ together with $\alpha\rightarrow
\alpha+\lambda$. When the coupling parameter $\eta\rightarrow0$ and
the graviton mass $m\rightarrow0$, our model reduces to the
generalized holographic superconductors of the Einstein's general
relativity (GR) investigated in \cite{MahapatraJHEP2014}.

Using the gauge symmetry to fix the phase $\alpha=0$, from the
action (\ref{Lagrangian}) we can obtain equations of motion for the
matter fields
\begin{eqnarray}
\frac{1}{\sqrt{-g}}\partial_{\mu}\left[\sqrt{-g}(g^{\mu\nu}+\eta
G^{\mu\nu})(\partial_{\nu}\Psi)\right]-m_{\Psi}^2\Psi-\frac{d\textrm{G}(\Psi)}{d\Psi}(g^{\mu\nu}+\eta
G^{\mu\nu})A_{\mu}A_{\nu}=0,\label{PsiUniversalEquationMotion}
\end{eqnarray}
\begin{eqnarray}
\frac{1}{\sqrt{-g}}\partial_{\mu}(\sqrt{-g}F^{\mu\nu})-2\textrm{G}(\Psi)(g^{\mu\nu}+\eta
G^{\mu\nu})A_{\mu}=0. \label{PhiUniversalEquationMotion}
\end{eqnarray}
Taking the ansatz of the matter fields as $\Psi=\Psi(r)$ and
$A=\Phi(r)dt$, under the solution (\ref{MassiveBH}), the above
equations of motion can be rewritten into
\begin{eqnarray}
\biggl[1+\eta(3 f+r f^{\prime})\biggr]\Psi''
&+&\biggl[\frac{4}{r}+\frac{f'}{f}+\eta\left(rf^{\prime \prime}+11f^{\prime}+\frac{r f'^{2}}{f}+\frac{12f}{r}\right)\biggr]\Psi'\nonumber \\
&+&\frac{1}{r^{2}f}\biggl\{-
m_{\Psi}^{2}\Psi+\Phi^{2}\biggl[\frac{1}{r^{2}f}+\eta(\frac{3}{r^{2}}+\frac{f^{\prime}}{rf})\biggr]\frac{d\textrm{G}(\Psi)}{d\Psi}\biggr\}=0,
\label{PsiEquationMotion}
\end{eqnarray}
\begin{eqnarray}
\Phi''+\frac{2}{r}\Phi'-\frac{2\textrm{G}(\Psi)}{r^{2}f}\biggl[1+\eta(3
f+r f^{\prime})\biggr]\Phi=0. \label{PhiEquationMotion}
\end{eqnarray}
Using the shooting method \cite{HartnollJHEP12}, we can solve Eqs.
(\ref{PsiEquationMotion}) and (\ref{PhiEquationMotion}) numerically
by doing integration from the horizon out to the infinity with the
appropriate boundary conditions for $\Psi$ and $\Phi$. At the event
horizon $r=r_{h}$, the regularity leads to the boundary conditions
\begin{equation}
\Phi(r_{h})=0,~~~~\Psi'(r_{h})=\frac{m_{\Psi}^2}{r_{h}^2f'(r_{h})[1+\eta
r_{h}f'(r_{h})]}\Psi(r_{h}). \label{Horizon}
\end{equation}
At the infinity $r\rightarrow\infty$, the solutions behave like
\begin{equation}
\Phi=\mu-\frac{\rho}{r},~~~~\Psi=\frac{\Psi_{-}}{r^{\Delta_{-}}} +
\frac{\Psi_{+}}{r^{\Delta_{+}}}, \label{Infinity}
\end{equation}
with the characteristic exponent
\begin{eqnarray}
\Delta_\pm=\frac{1}{2}\left(3\pm\sqrt{9+\frac{4m_{\Psi}^{2}}{1+3
\eta}}\right),\label{Exponent}
\end{eqnarray}
where $\mu$ and $\rho$ are interpreted as the chemical potential and
charge density in the dual field theory, respectively. According to
the AdS/CFT correspondence, provided $\Delta_{-}$ is larger than the
unitarity bound, both $\Psi_{-}$ and $\Psi_{+}$ can be normalizable
and they can be used to define operators in the dual field theory
$\Psi_{-}=\langle O_{-}\rangle$ and $\Psi_{+}=\langle O_{+}\rangle$,
respectively. Considering that we focus on the condensate for the
operator $\langle O_{+}\rangle$, we will impose boundary condition
$\Psi_{-}=0$ in this work.

\section{Scalar condensation and phase transition}

In this section, we will solve the system of coupled differential
equations (\ref{PsiEquationMotion}) and (\ref{PhiEquationMotion})
numerically and then discuss the scalar condensation and phase
transition of these generalized superconductors. For concreteness,
we will set $m_{\Psi}^2=-2$ since the choices of the scalar field
mass will not qualitatively modify our results. Following
\cite{MahapatraJHEP2014,ChengEPJC2018}, we consider a particular
form of $\textrm{G}(\Psi)$ for simplicity, i.e.,
\begin{equation}
\textrm{G}(\Psi)=\Psi^{2}+\xi\Psi^{8},
\end{equation}
where $\xi$ is the model parameter. In order to compare with the
results given in Refs. \cite{MahapatraJHEP2014,ChengEPJC2018}, we
will concentrate on the cases for the model parameters  $\xi=0.2$
and $\xi=0.5$ in our discussion. For the convenience of numerics, we
will set $c_{1}=r_{h}$ and $c_{2}=-r_{h}^{2}/2$, and choose the
range of graviton mass parameters $0\leq m\leq 1.2$, just as in
\cite{HuPRD2016}. It is interesting to note that, from the equations
of motion (\ref{PsiEquationMotion}) and (\ref{PhiEquationMotion}),
there exist the useful scaling symmetries and the transformation of
the relevant quantities
\begin{eqnarray}
&&r\rightarrow \lambda r,~~~(t,x,y)\rightarrow
\lambda^{-1}(t,x,y),~~~c_{1}\rightarrow\lambda
c_{1},~~~c_{2}\rightarrow\lambda^{2} c_{2},
\nonumber \\
&&\Psi\rightarrow\Psi,~~~\Phi\rightarrow
\lambda\Phi,~~~(T,\mu)\rightarrow
\lambda(T,\mu),~~~\rho\rightarrow\lambda^{2}\rho,~~~\Psi_{\pm}\rightarrow
\lambda^{\Delta_{\pm}}\Psi_{\pm},
\end{eqnarray}
where $\lambda$ is a real positive number. We will use the scaling
symmetries to set $r_{h}=1$ when performing numerical calculations.

On the other hand, we have to calculate the grand potential
$\Omega=-\mathcal{T}\mathcal{S}_{os}$ of the bound state in order to
obtain the important information about the phase transition of the
system, where $\mathcal{S}_{os}$ is the Euclidean on-shell action.
Working in the grand canonical ensemble, we have
\begin{eqnarray}
\mathcal{S}_{os}&=&\int dtdxdydr \biggl\{-\frac{1}{2}\partial_{\mu}(\sqrt{-g}F^{\mu\nu}A_{\nu})+\frac{1}{2}A_{\nu}\partial_{\mu}(\sqrt{-g}F^{\mu\nu})-\frac{1}{2}\partial_{\mu}[\sqrt{-g}\Psi(g^{\mu\nu}+\eta G^{\mu\nu})(\partial_{\nu}\Psi)] \nonumber \\
&&~~~~~~~~~~~~~~~~~+\frac{1}{2}\Psi\partial_{\mu}[\sqrt{-g}(g^{\mu\nu}+\eta
G^{\mu\nu})(\partial_{\nu}\Psi)]-\frac{1}{2}m_{\Psi}^2\Psi^2-\textrm{G}(\Psi)(g^{\mu\nu}+\eta G^{\mu\nu})A_{\mu}A_{\nu}\biggr\} \nonumber \\
&=&\frac{V}{2\mathcal{T}}\biggl\{r^{2}\Phi
\Phi^{\prime}|_{r\rightarrow
\infty}-r^{4}f[1+\eta(3f+rf^{\prime})]\Psi\Psi^{\prime}|_{r\rightarrow\infty}
+\int^{\infty}_{r_{h}}dr\sqrt{-g}\biggl[\Psi\frac{d\textrm{G}(\Psi)}{d\Psi}(g^{\mu\nu}+\eta G^{\mu\nu})A_{\mu}A_{\nu}\biggr] \biggr\} \nonumber \\
&=&\frac{V}{2\mathcal{T}}\biggl[\mu\rho-\int^{\infty}_{r_{h}}\Psi\frac{d\textrm{G}(\Psi)}{d\Psi}\frac{1+\eta(3f+rf^{\prime})}{f}\Phi^{2}dr\biggr],
\end{eqnarray}
where we have used the integration $\int
dtdxdy=\frac{V}{\mathcal{T}}$. Thus, we can express the grand
potential in the superconducting phase as
\begin{eqnarray}
\frac{\Omega_{S}}{V}=-\frac{\mathcal{T}\mathcal{S}_{os}}{V}=-\frac{1}{2}\mu\rho
+\frac{1}{2}\int^{\infty}_{r_{h}}\Psi\frac{d\textrm{G}(\Psi)}{d\Psi}\frac{1+\eta(3f+rf^{\prime})}{f}\Phi^{2}dr.
\end{eqnarray}
In the normal phase, i.e., $\Psi=0$, we can easily get the grand
potential
\begin{eqnarray}
\frac{\Omega_{N}}{V}=-\frac{1}{2}\mu\rho.
\end{eqnarray}

\subsection{Effect of the graviton mass on the scalar condensation
and phase transition}

\begin{figure}[ht]
\includegraphics[scale=0.6]{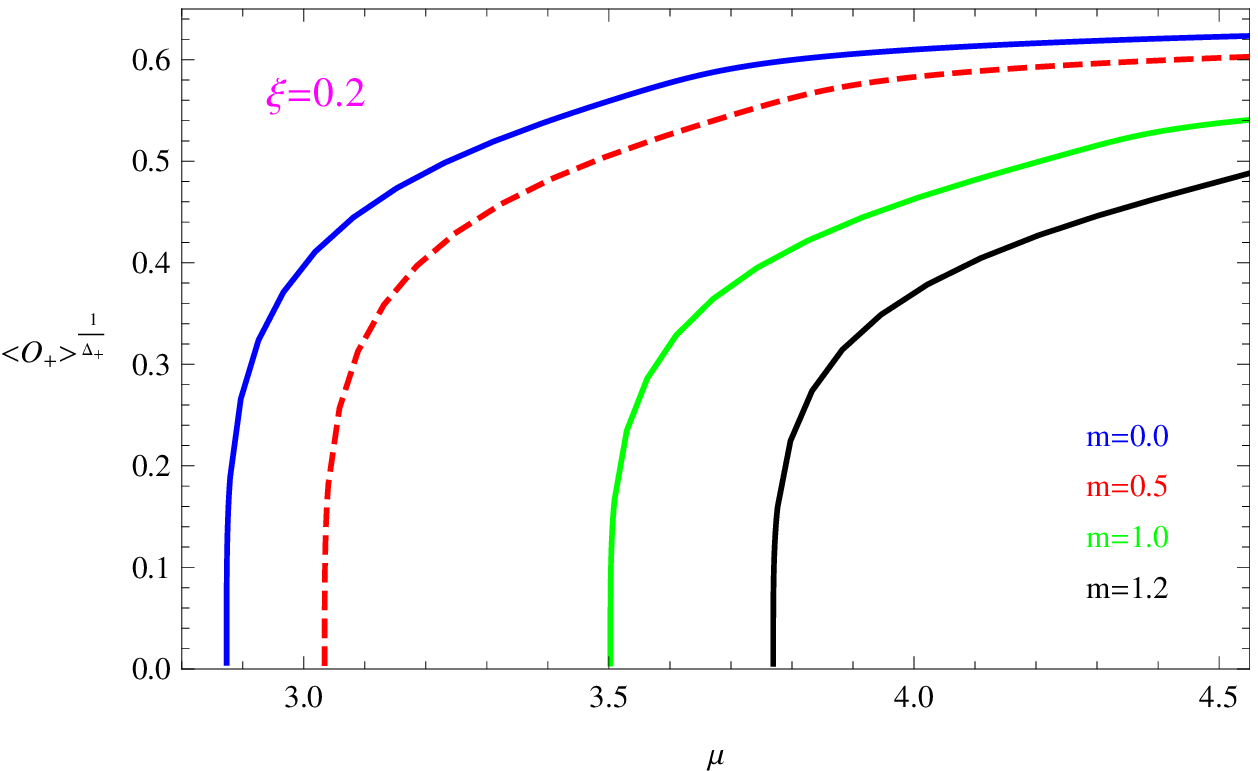}\hspace{0.2cm}%
\includegraphics[scale=0.6]{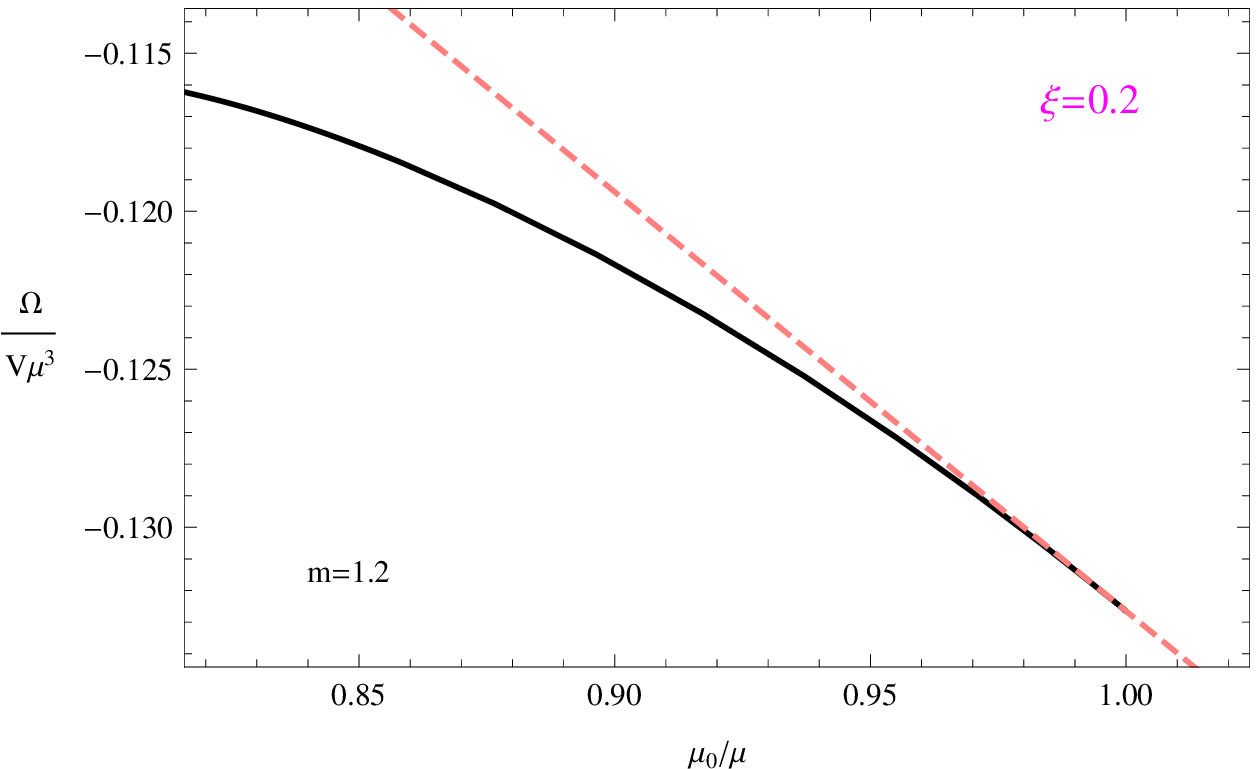}\\ \vspace{0.0cm}
\includegraphics[scale=0.6]{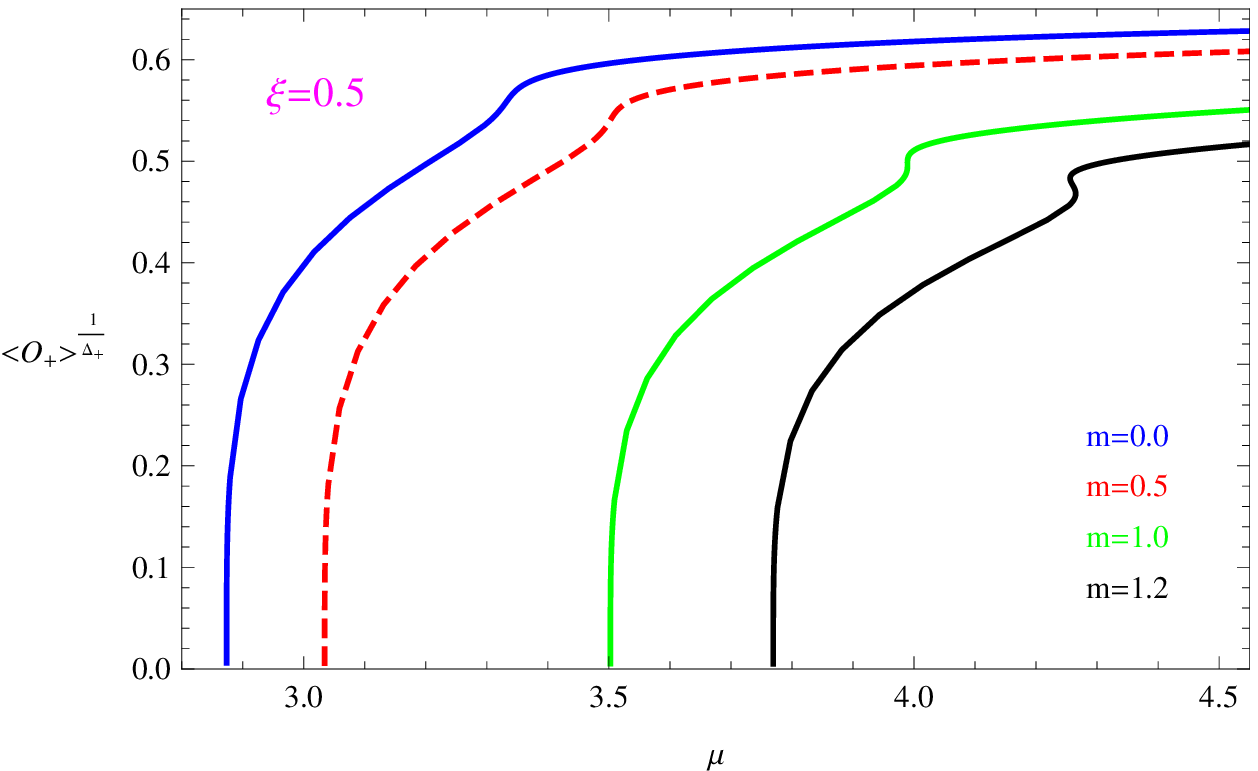}\hspace{0.2cm}%
\includegraphics[scale=0.6]{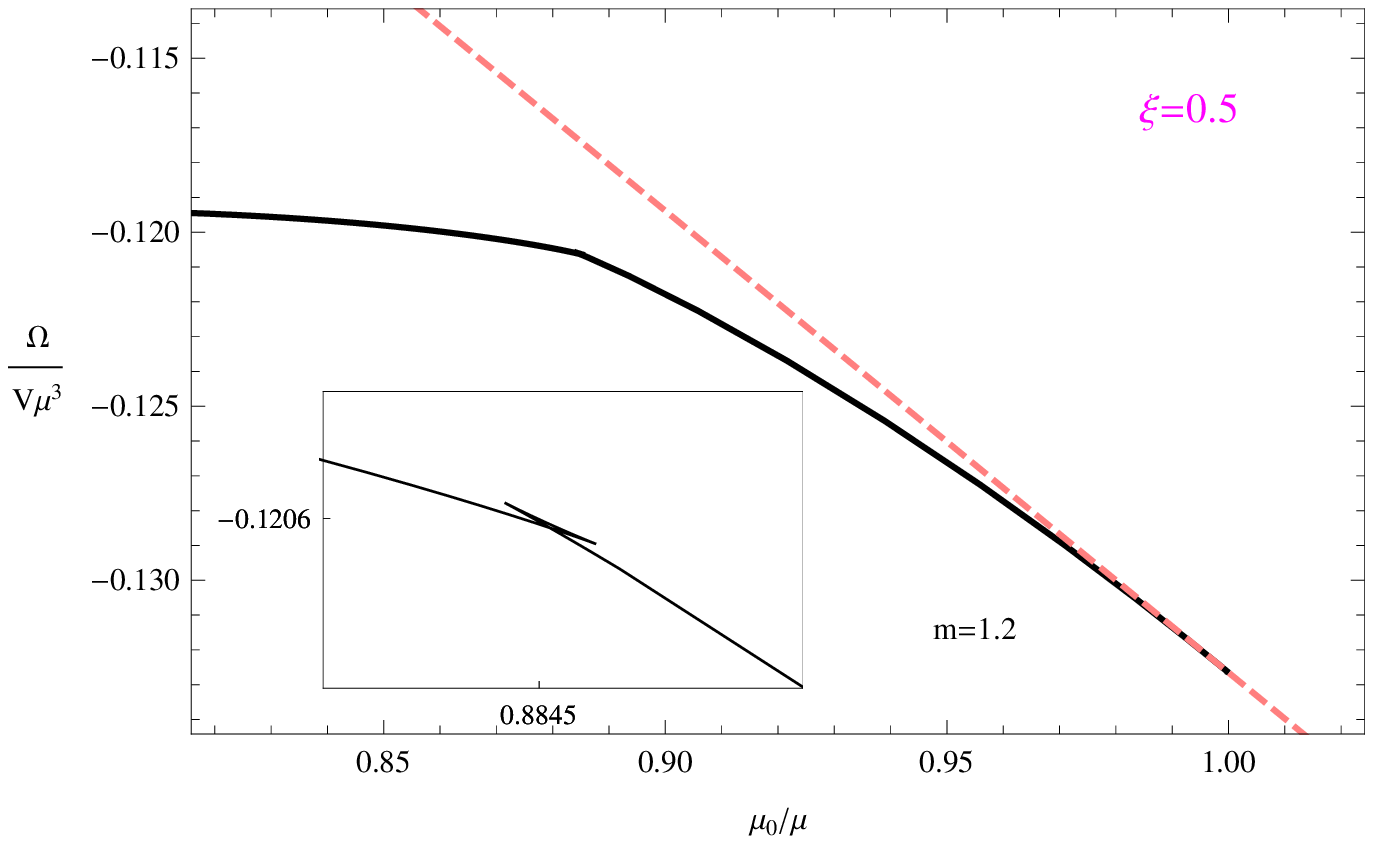}\\ \vspace{0.0cm}
\caption{\label{mpsi2gravCondensate} (Color online) The condensate
and grand potential as a function of the chemical potential with the
fixed coefficients $\xi=0.2$ (top) and $\xi=0.5$ (bottom) for
different values of the graviton mass $m$ in the case of the
coupling parameter $\eta=0.0$.  For the left two panels, the four
lines in each panel from left to right correspond to increasing
graviton mass, i.e., $m=0.0$ (blue), $0.5$ (red dotted), $1.0$
(green) and $1.2$ (black), respectively. For the right two panels,
the two lines in each panel correspond to the superconducting phase
(black) and normal phase (pink dotted) in the case of $m=1.2$,
respectively. }
\end{figure}

In Fig. \ref{mpsi2gravCondensate}, we plot the condensate and
corresponding grand potential as a function of the chemical
potential with the fixed coefficient $\xi$ for different values of
the graviton mass $m$ in the case of the coupling parameter
$\eta=0.0$ by solving the equations of motion
(\ref{PsiEquationMotion}) and (\ref{PhiEquationMotion}) numerically.
Obviously, we can see clearly from left two panels that, for a fixed
model parameter $\xi$, the critical chemical potential $\mu_{c}$
increases with the increase of the graviton mass parameter, i.e.,
the critical temperature $T_{c}$ decreases as $m$ increases, which
indicates that the larger graviton mass parameters will hinder the
formation of the condensation. This agrees well with the findings in
the first holographic Josephson junction from the massive gravity
introduced in Ref. \cite{HuPRD2016}, i.e., the graviton mass
parameter and the doping have a similar effect as the phase
transition from superconductivity to a normal metal.

What is more interesting is the rich phase structure of this system.
In the case of $\xi=0$ or small $\xi$, for example $\xi=0.2$, the
transition is of the second order and the condensate approaches zero
as $\langle O_{2}\rangle\sim (\mu-\mu_{c})^{1/2}$ for all values of
$m$ considered here, which is supported by the behavior of the grand
potential just as shown in the top-right panel for $m=1.2$. Note
that this conclusion still holds in the cases of $m\leq1.0$ with
$\xi=0.5$. However, the story is obviously different if we consider
the case of $m=1.2$ in the bottom two panels of Fig.
\ref{mpsi2gravCondensate}. We find that, as the chemical potential
is increased, i.e., the temperature is lowered, the condensate
$\langle O_{+}\rangle$ becomes multivalued, and there is at first a
second order phase transition from the normal to a superconducting
phase and then a first order phase transition between two
superconducting phases, which is the Cave of Winds \cite{Arean2010}
and consistent with the behavior of the grand potential in the
bottom-right panel. Thus, we conclude that the combination of the
graviton mass and the model parameter provides richer physics in the
phase transition, and the larger graviton mass makes it easier for
the emergence of the Cave of Winds. It would be very interesting to
see if there are any real systems with doping in the condensed
matter which show this kind of Cave of Winds structure.

\subsection{Effect of the Einstein tensor on the scalar condensation
and phase transition}

\begin{figure}[ht]
\includegraphics[scale=0.6]{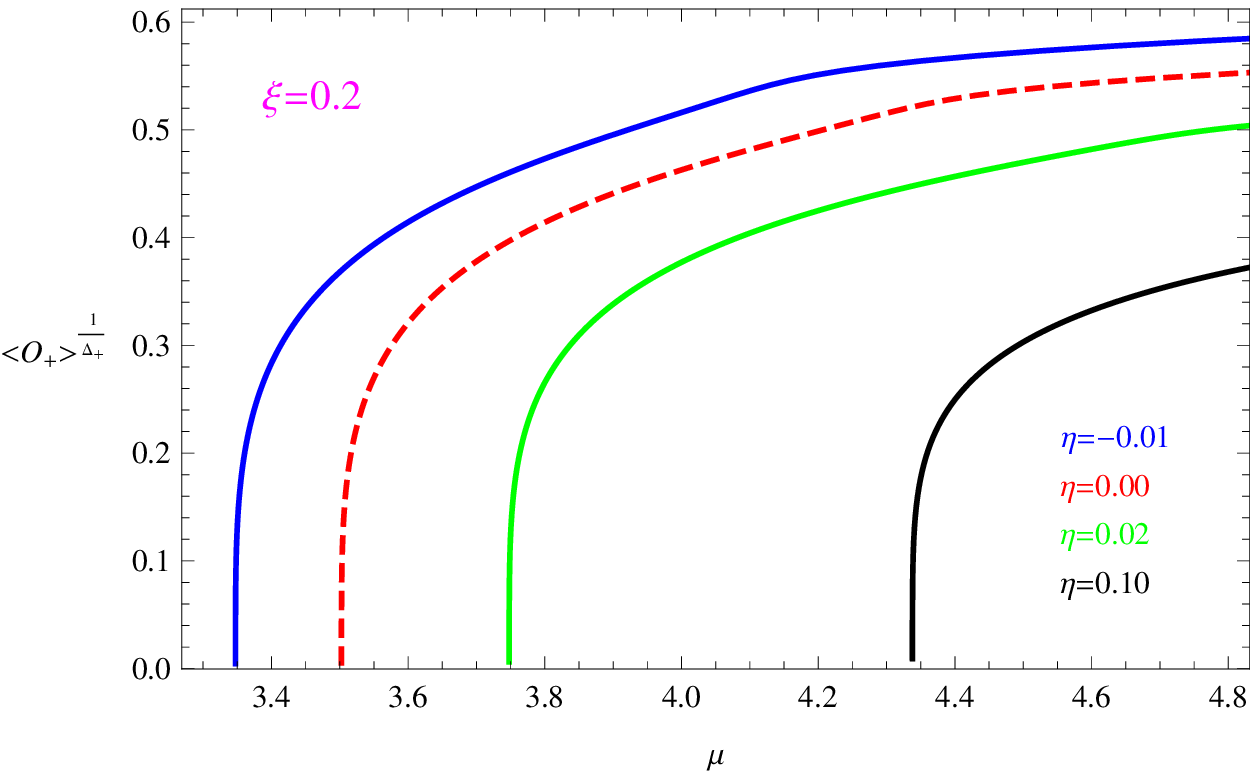}\hspace{0.2cm}%
\includegraphics[scale=0.6]{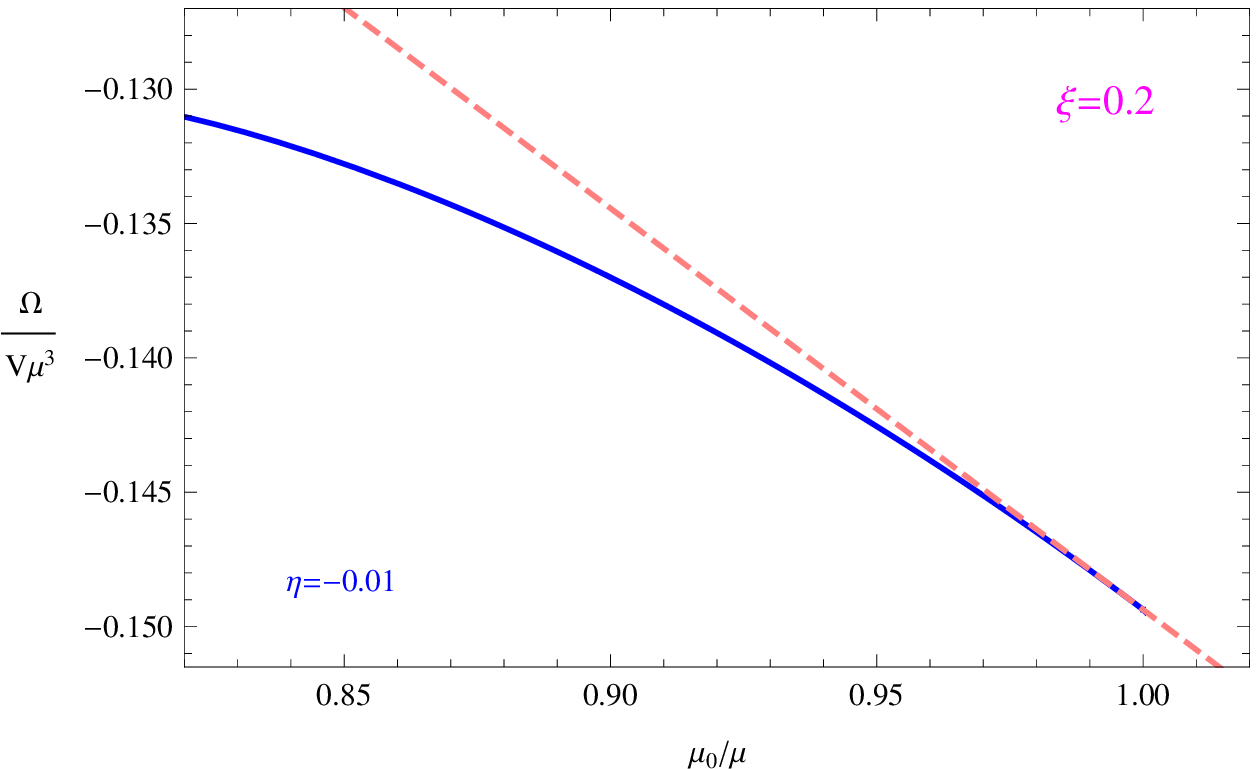}\\ \vspace{0.0cm}
\includegraphics[scale=0.6]{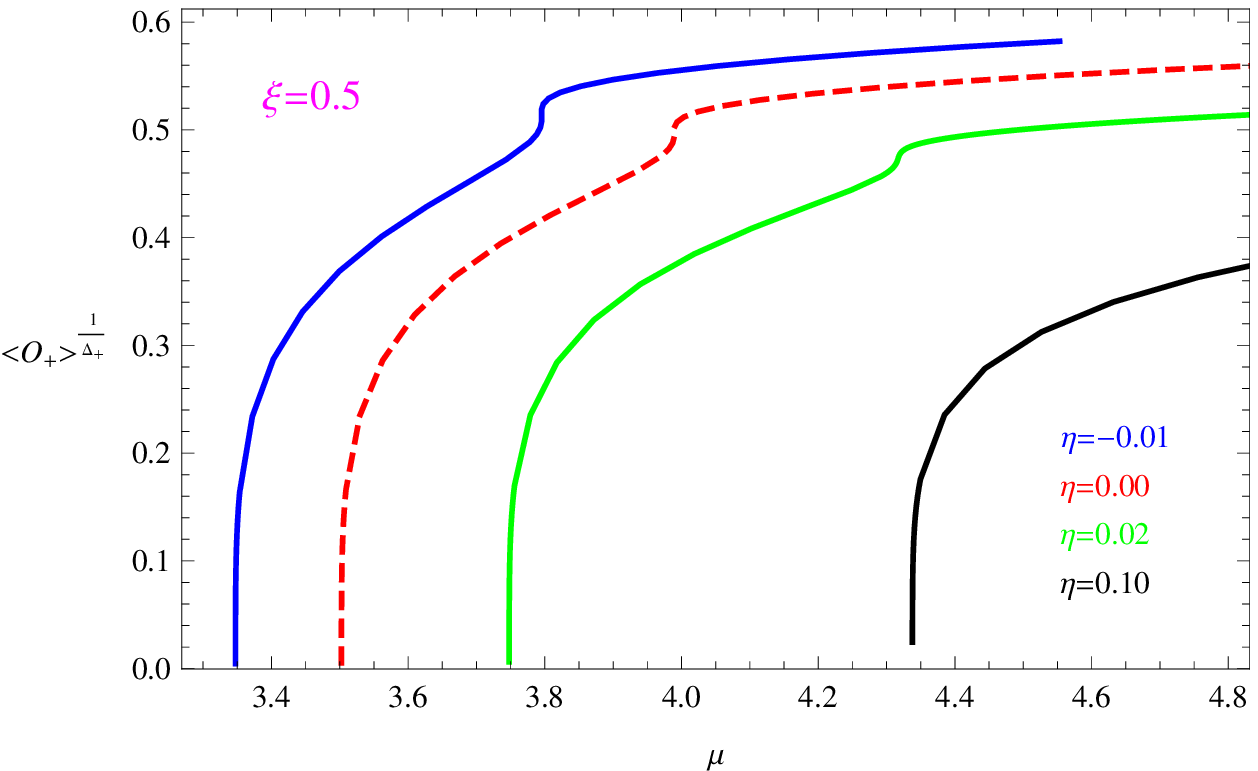}\hspace{0.2cm}%
\includegraphics[scale=0.6]{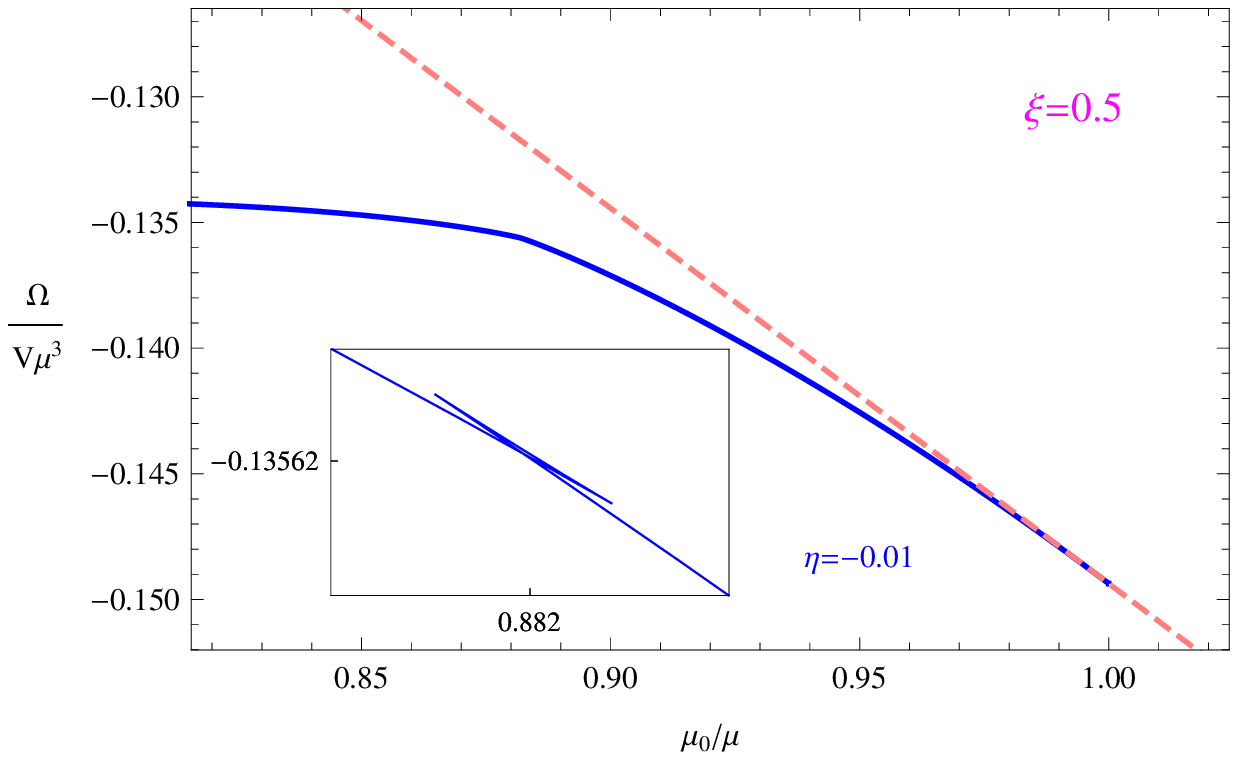}\\ \vspace{0.0cm}
\caption{\label{mpsi2eatCondensate} (Color online) The condensate
and grand potential as a function of the chemical potential with the
fixed coefficient $\xi=0.2$ (top) and $\xi=0.5$ (bottom) for
different values of the coupling parameter $\eta$ in the case of the
graviton mass $m=1.0$. For the left two panels, the four lines in
each panel from top to bottom correspond to increasing coupling
parameter, i.e., $\eta=-0.01$ (blue), $0.00$ (red dotted), $0.02$
(green) and $0.10$ (black), respectively. For the right two panels,
the two lines in each panel correspond to the superconducting phase
(blue) and normal phase (pink dotted) in the case of $\eta=-0.01$,
respectively. }
\end{figure}

We now move on to the effect of the Einstein tensor on the scalar
condensation and phase transition. Fixing the graviton mass by
$m=1.0$, we exhibit the condensate and grand potential as a function
of the chemical potential with the coefficient $\xi=0.2$ (top) and
$\xi=0.5$ (bottom) for different values of the coupling parameter
$\eta$ in Fig. \ref{mpsi2eatCondensate}. From left two panels, we
observe that, similar to the effect of the graviton mass parameter,
the critical chemical potential $\mu_{c}$ increases as the coupling
parameter $\eta$ increases for a fixed model parameter $\xi$, which
implies that the Einstein tensor will hinder the condensate of the
scalar field. This can be used to back up the observation obtained
in Ref. \cite{KuangE2016} and suggests that this derivative coupling
in the gravity bulk can have a dual interpretation on the boundary
corresponding to impurities concentrations in a real material.

Focusing on the phase structure of this system, we find that the
transition is of the second order and the condensate approaches zero
as $\langle O_{2}\rangle\sim (\mu-\mu_{c})^{1/2}$ for all values of
$\eta$ considered here in the case of $\xi=0$ or small $\xi$, for
example $\xi=0.2$, which is in good agreement with the behavior of
the grand potential just as shown in the top-right panel for
$\eta=-0.01$. Interestingly, in the case of large $\xi$, for example
$\xi=0.5$, the conclusion still holds for the large coupling
parameter, for example $\eta\geq0.00$ considered here. But in the
case of $\eta=-0.01$ with $\xi=0.5$, we show that, as the chemical
potential is increased, the condensate $\langle O_{+}\rangle$
becomes multivalued and the second-order transition occurs before
the first-order transition to a new superconducting phase, which is
supported by the behavior of the grand potential as shown in the
bottom-right panel. Obviously, the smaller the coupling parameter
the easier it is for the Cave of Winds to emerge, which is different
from the effect of the graviton mass parameter. Of course, it is
worthwhile to see if there are any real systems with impurities in
the condensed matter which present this kind of the Cave of Winds
structure.

\section{Negative refractive index}

In previous section, we have revealed the effects of the graviton
mass and Einstein tensor on the scalar condensation and phase
transition, i.e., the graviton mass and Einstein tensor have the
similar effect on the critical chemical potential but completely
different effect on the emergence of the Cave of Winds. Now we are
going to investigate the effects of the graviton mass and Einstein
tensor on the optical properties of the generalized superconductors,
especially on the negative refraction in these holographic systems.

Before proceeding, we first briefly review the electric permittivity
$\epsilon(\omega)$, effective magnetic permeability $\mu(\omega)$
and their dependence on the retarded correlators in the linear
response theory investigated by Amariti \emph{et al.}
\cite{AmaritiFMP2011}. Expanding the transverse part of the
dielectric tensor for the small wave vector $k$ as
\begin{eqnarray}
\epsilon_{T}(\omega,k)=\epsilon(\omega)+\frac{k^{2}}{\omega^{2}}\left[1-\frac{1}{\mu(\omega)}\right]+O(k^{4}),
\end{eqnarray}
one can obtain the refractive index for the isotropic media
\cite{AmaritiFMP2011}
\begin{eqnarray}
n^{2}(\omega)=\frac{k^{2}}{\omega^{2}}=\epsilon(\omega)\mu(\omega).
\label{RefractiveIndex}
\end{eqnarray}
With the form of $\epsilon_{T}(\omega,k)$ specified by the
transverse part of the retarded current-current correlator
$G_{T}(\omega,k)$ \cite{LandauL,DresselG}, i.e.,
\begin{eqnarray}
\epsilon_{T}(\omega,k)=1+\frac{4\pi}{\omega^{2}}C_{em}^{2}G_{T}(\omega,k),
\end{eqnarray}
and the expanded form of the retarded correlator in $k$
\begin{eqnarray}
G_{T}\left( \omega,k \right) =G_{T}^{0}\left( \omega \right) +k^{2}
G_{T}^{2}\left( \omega \right) +O(k^{4}), \label{GTExpand}
\end{eqnarray}
the electric permittivity and effective magnetic permeability can be
expressed as
\begin{eqnarray}
&&\epsilon \left( \omega \right) = 1+\frac{4\pi }{\omega ^{2}}
C_{em}^{2}G_{T}^{0}\left( \omega \right),
\nonumber \\
&&\mu\left( \omega \right) = \frac{1}{1- 4\pi
C_{em}^{2}G_{T}^{2}\left( \omega \right) }, \label{epsmu}
\end{eqnarray}
where $C_{em}$ is the electromagnetic coupling constant which will
be set to unity when performing numerical calculations, and
$G_{T}^{0}\left( \omega \right)$ and $G_{T}^{2}\left( \omega
\right)$ are the expansion coefficients of the retarded correlators
\cite{SonStarinets}, respectively.

It should be noted that, when taking the dissipation of the system
into account, the electric permittivity $\epsilon$, magnetic
permeability $\mu$ and refractive index $n$ are generally complex
functions of the frequency $\omega$. In Ref. \cite{McCallLW}, McCall
\emph{et al.} derived the condition for the negative refraction in
the dissipative medium and pointed out that it is not necessary for
real parts $Re(\epsilon)<0$ and $Re(\mu)<0$ simultaneously. Later,
Depine and Lakhtakia found that, for the dissipative medium, the
sign of the refractive index coincide with the sign of the DL index
\cite{DepineLakhtakia}
\begin{eqnarray}\label{DLIndex}
n _{DL}= Re[\epsilon(\omega)]|\mu(\omega)| +Re[\mu(\omega)]|
\epsilon(\omega)|,
\end{eqnarray}
where $n _{DL}<0$ indicates that the phase velocity in the medium is
opposite to the direction of the energy flow, i.e., the system has a
negative index of refraction. As a simple, convenient, and widely
adopted criterion, we will use the DL index $n _{DL}$ to check if
the medium has positive or negative refraction in this work.

In order to compute the refractive index of our holographic systems,
we will use the AdS/CFT correspondence to derive the expression of
$G_{T}^{0}(\omega)$ and $G_{T}^{2}(\omega)$. Assuming that the
perturbed Maxwell field has a form $\delta A_{x}= A_{x}(r) e^{-i
\omega t + iky}$, we get the equation of motion
\begin{eqnarray}
A_{x}''+\biggl(\frac{2}{r}+\frac{f'}{f} \biggr)A_{x}'
+\frac{1}{r^{2}f}
\biggl\{\frac{\omega^{2}}{r^{2}f}-\frac{k^{2}}{r^{2}}-\textrm{G}(\Psi)
\biggl[ 2+ \eta(6 f + 6 r f'+r^{2} f'')\biggr]\biggr\} A_{x}=0.
\label{AxEM}
\end{eqnarray}
Taking the same series expansion of $A_{x}$ as in Eq.
(\ref{GTExpand}) for $G_{T}(\omega,k)$, i.e.,
\begin{eqnarray}
A_{x}(r)=A_{x0}(r)+k^{2}A_{x2}(r)+O(k^{4}), \label{AxExpand}
\end{eqnarray}
from Eq. (\ref{AxEM}) we arrive at the equations of motion for
$A_{x0}$ and $A_{x2}$
\begin{eqnarray}
&&A_{x0}''+\biggl(\frac{2}{r}+\frac{f'}{f} \biggr)A_{x0}'
+\frac{1}{r^{2}f} \biggl\{\frac{\omega^{2}}{r^{2}f}-\textrm{G}(\Psi)
\biggl[ 2+ \eta(6 f + 6 r f'+r^{2} f'')\biggr]\biggr\} A_{x0}=0, \label{Ax0} \\
&&A_{x2}''+\biggl(\frac{2}{r}+\frac{f'}{f} \biggr)A_{x2}'
+\frac{1}{r^{2}f} \biggl\{\frac{\omega^{2}}{r^{2}f}-\textrm{G}(\Psi)
\biggl[ 2+ \eta(6 f + 6 r f'+r^{2} f'')\biggr]\biggr\}
A_{x2}-\frac{A_{x0}}{r^{4}f}=0, \label{Ax2}
\end{eqnarray}
which can be numerically solved by the ingoing wave boundary
conditions near the horizon
\begin{eqnarray}
A_{x0}\propto f^{-\frac{i\omega}{(3+m^{2}/2)r_{h}}},~~~A_{x2}\propto
f^{-\frac{i\omega}{(3+m^{2}/2)r_{h}}},
\end{eqnarray}
and the behaviors near the asymptotic AdS boundary
\begin{equation}
A_{x0} = A_{x0}^{(0)} + \frac{A_{x0}^{(1)}}{r},~~~A_{x2} =
A_{x2}^{(0)} + \frac{A_{x2}^{(1)}}{r}.
\end{equation}
Thus, according to the AdS/CFT dictionary, we obtain the retarded
correlators
\begin{equation}
G_{T}^0 (\omega)=\frac{A_{x0}^{(1)}}{A_{x0}^{(0)}},~~~G_{T}^2
(\omega)=\frac{A_{x0}^{(1)}}{A_{x0}^{(0)}}
\biggl[\frac{A_{x2}^{(1)}}{A_{x0}^{(1)}} -
\frac{A_{x2}^{(0)}}{A_{x0}^{(0)}} \biggr]\label{GT0GT2},
\end{equation}
which can be used to calculate $\epsilon(\omega)$ and $\mu(\omega)$
in Eq. (\ref{epsmu}) and $n _{DL}$ in Eq. (\ref{DLIndex}).

\subsection{Effect of the graviton mass on the negative refractive index}

\begin{figure}[ht]
\includegraphics[scale=0.6]{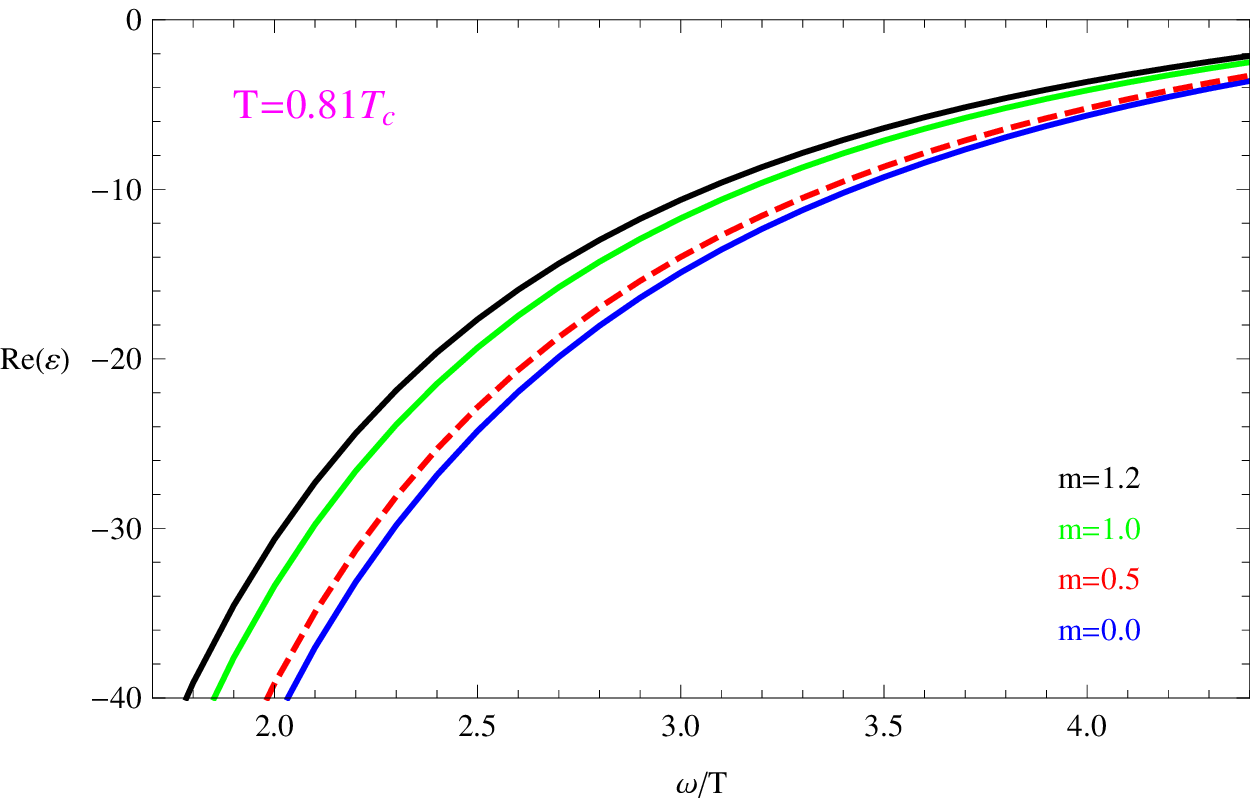}\hspace{0.2cm}%
\includegraphics[scale=0.6]{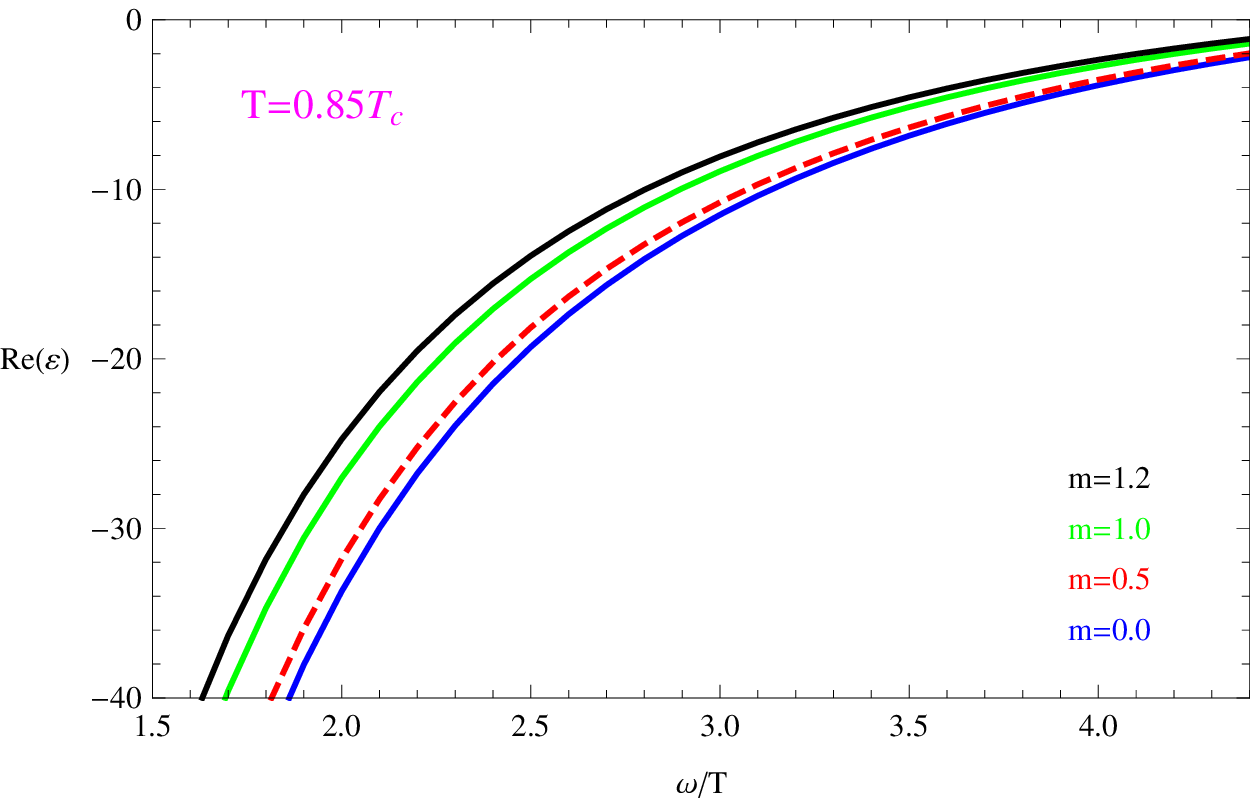}\\ \vspace{0.0cm}
\includegraphics[scale=0.6]{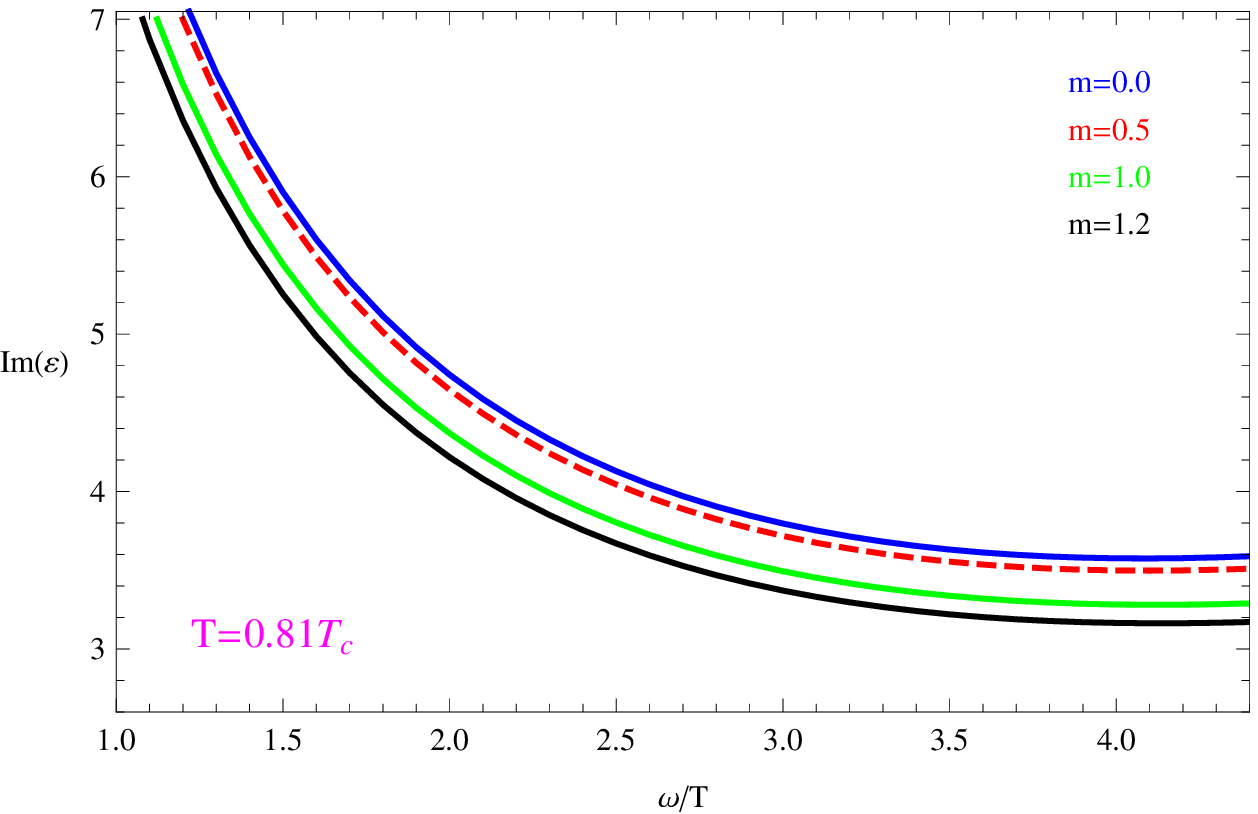}\hspace{0.2cm}%
\includegraphics[scale=0.6]{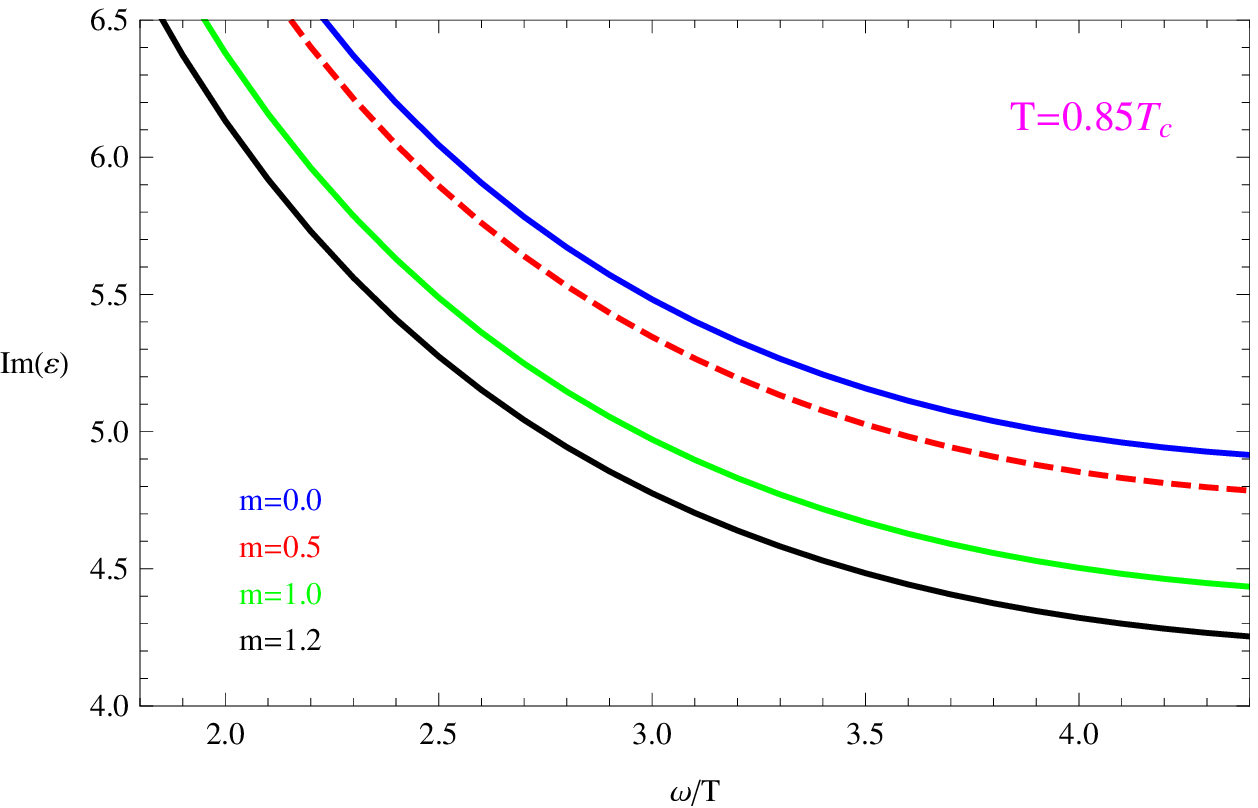}\\ \vspace{0.0cm}
\caption{\label{gravitonmassPermittivity} (Color online) Real (top)
and imaginary (bottom) parts of the permittivity $\epsilon$ as a
function of $\omega/T$ with the fixed temperatures $T=0.81T_{c}$
(left) and $T=0.85T_{c}$ (right) for different values of the
graviton mass parameter, i.e., $m=0.0$ (blue), $0.5$ (red dotted),
$1.0$ (green) and $1.2$ (black), respectively. We have set the
coupling parameter $\eta=0$ in the numerical computation. }
\end{figure}

\begin{figure}[ht]
\includegraphics[scale=0.6]{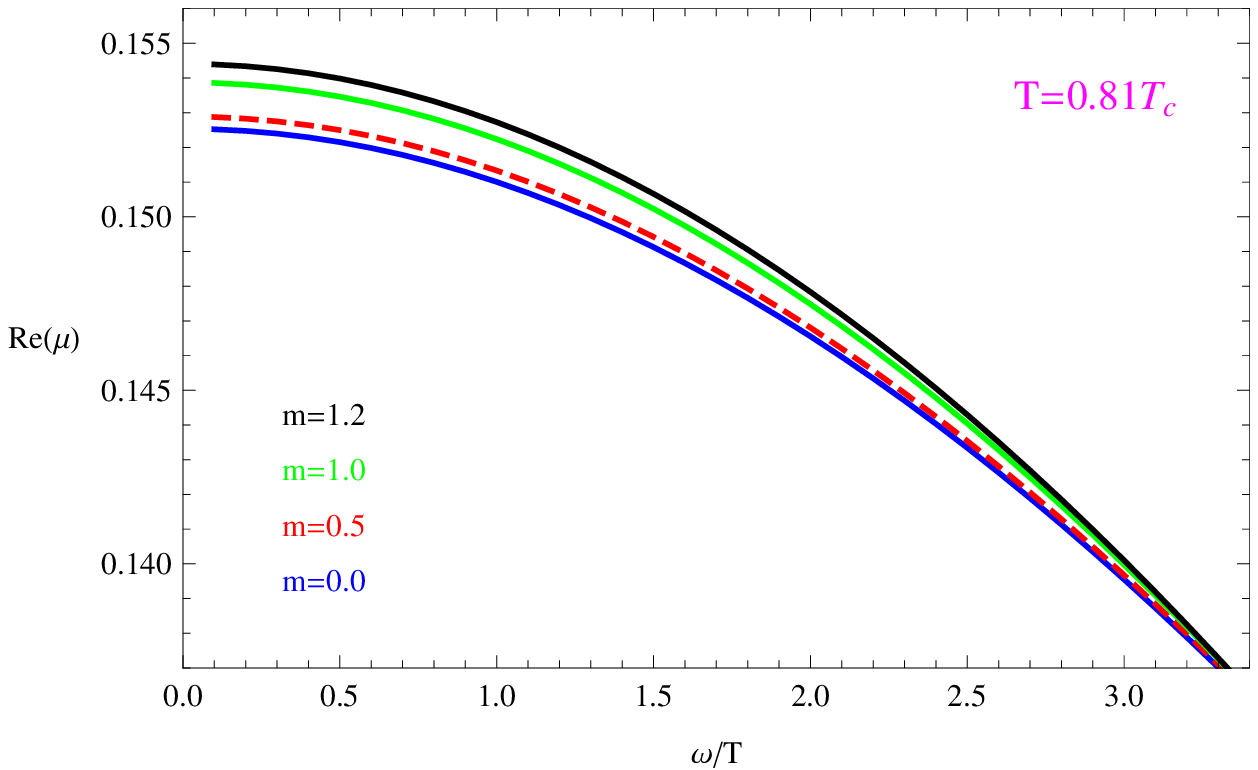}\hspace{0.2cm}%
\includegraphics[scale=0.6]{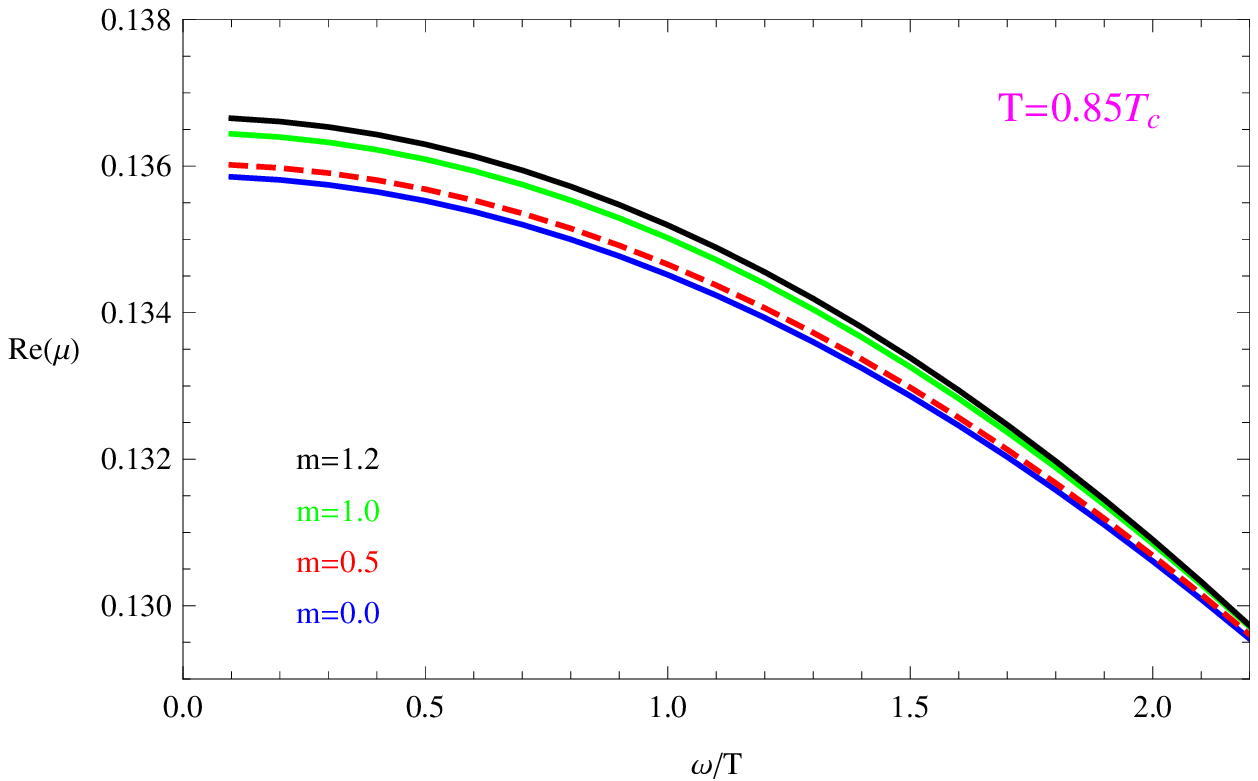}\\ \vspace{0.0cm}
\includegraphics[scale=0.6]{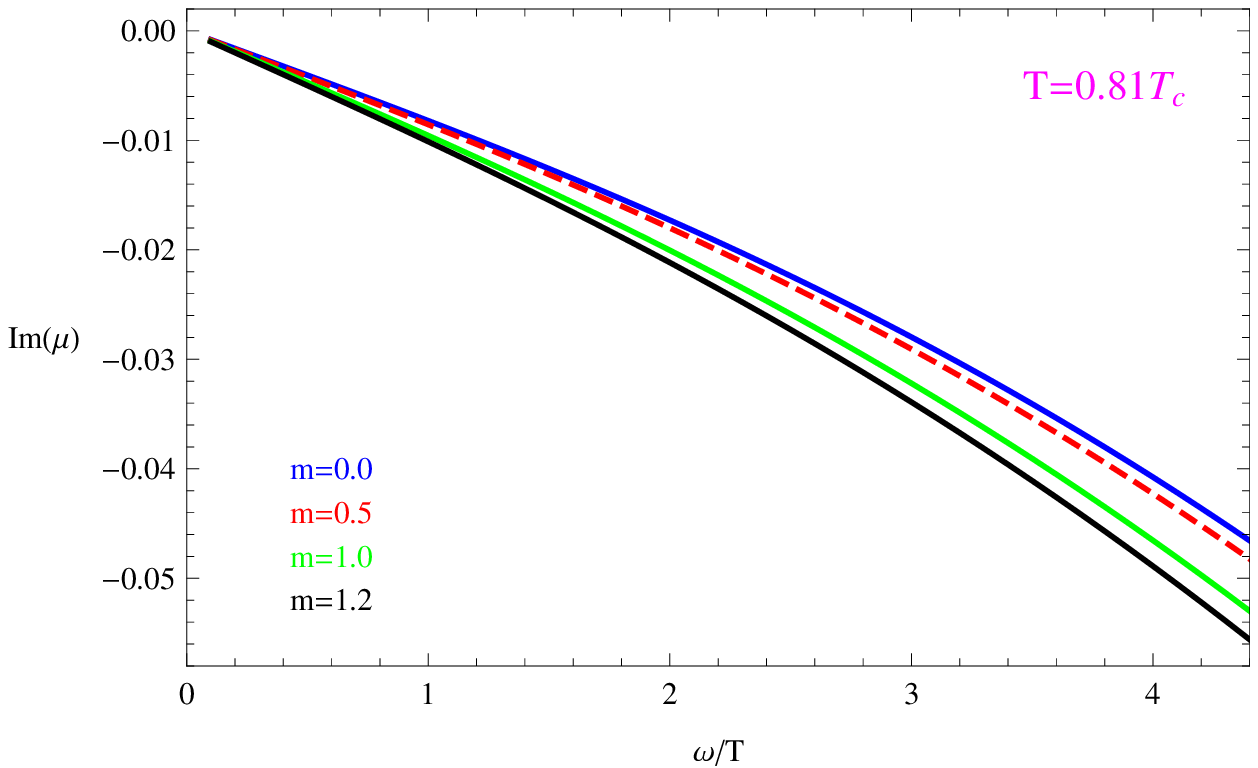}\hspace{0.2cm}%
\includegraphics[scale=0.6]{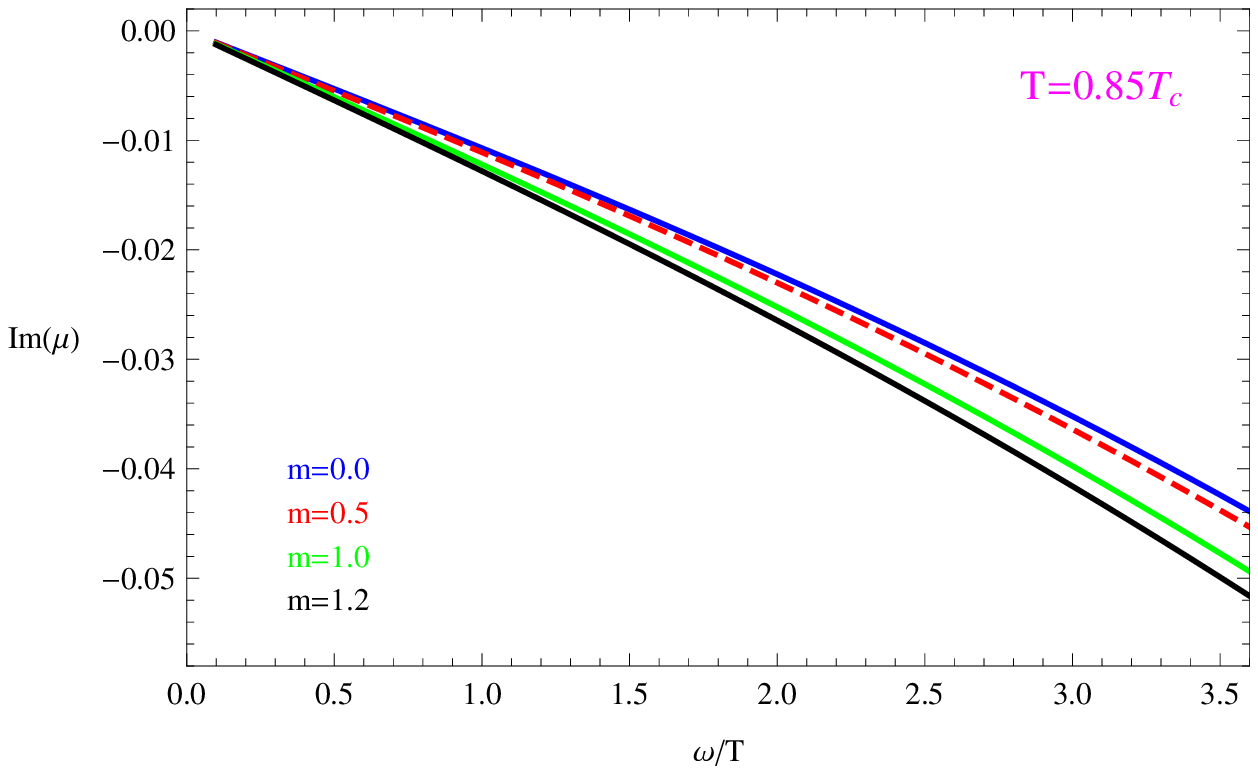}\\ \vspace{0.0cm}
\caption{\label{gravitonmassPermeability} (Color online) Real (top)
and imaginary (bottom) parts of the permeability $\mu$ as a function
of $\omega/T$ with the fixed temperatures $T=0.81T_{c}$ (left) and
$T=0.85T_{c}$ (right) for different values of the graviton mass
parameter, i.e., $m=0.0$ (blue), $0.5$ (red dotted), $1.0$ (green)
and $1.2$ (black), respectively. We have set the coupling parameter
$\eta=0$ in the numerical computation. }
\end{figure}

Let us firstly consider the effect of the graviton mass $m$ on the
negative refraction. Setting the coupling parameter $\eta=0$, in
Figs. \ref{gravitonmassPermittivity} and
\ref{gravitonmassPermeability} we plot the real and imaginary parts
of the permittivity $\epsilon$ and permeability $\mu$ as a function
of $\omega/T$ for different values of the graviton mass parameter
$m$. Since the system has negative refractive index in the range
$0.81T_{c}\leq T\leq0.85T_{c}$, we will fix the temperatures by
$T=0.81T_{c}$ and $T=0.85T_{c}$ in the numerical computation. From
the figures, we find that, for the fixed temperatures $T$ and all
values of $m$ considered here, Re$(\epsilon)$ and Im$(\mu)$ are
always negative but Im$(\epsilon)$ and Re$(\mu)$ are positive at low
frequencies. Just as pointed out in Ref. \cite{AmaritiFMP2011}, the
refraction may be negative when Re$(\epsilon)$ and Re$(\mu)$ are not
simultaneously negative because of the presence of the imaginary
parts. It should be noted that Im$(\mu)<0$ could imply some problem
in the $\epsilon-\mu$ approach
\cite{AmaritiFMS2011,MahapatraJHEP2014,DeyMT2014}, although Markel
argued that Im$(\mu)$ can in fact be negative for diamagnetic
materials \cite{Markel}. We will leave this point for further study
since this delicate issue goes beyond the scope of our paper.

\begin{figure}[ht]
\includegraphics[scale=0.6]{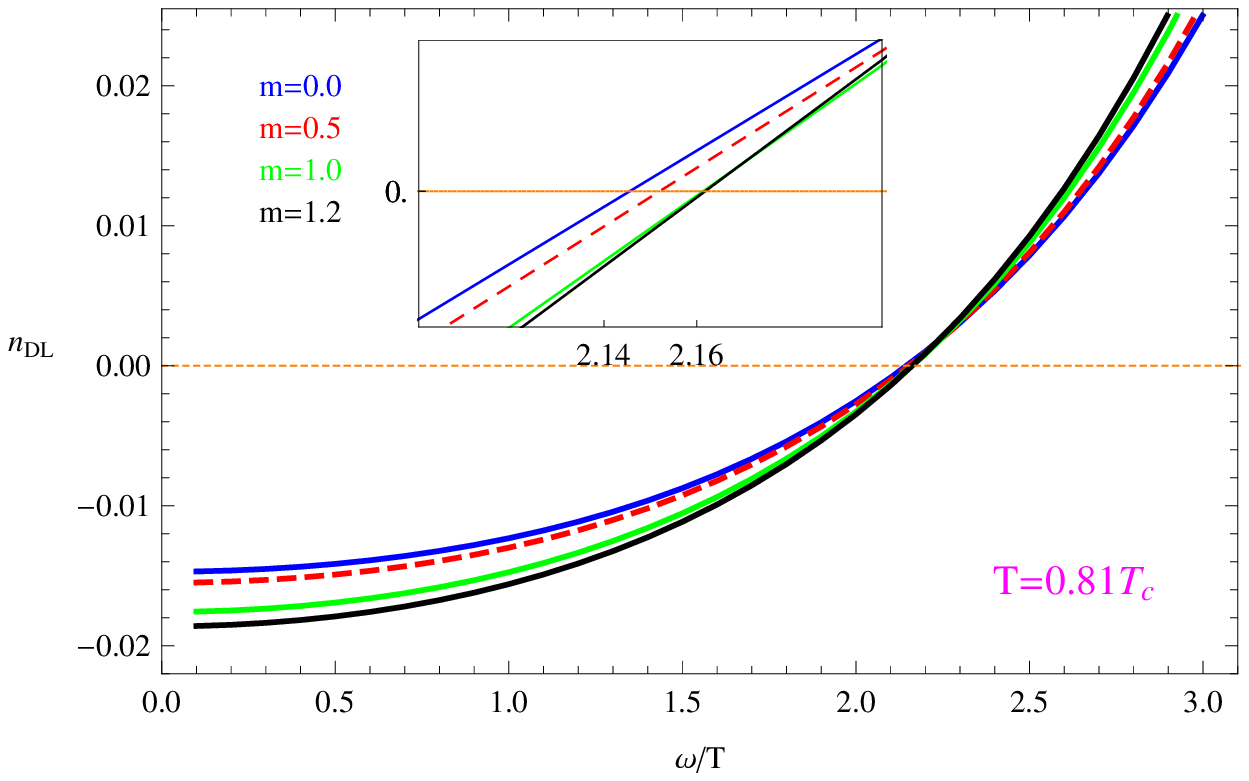}\hspace{0.2cm}%
\includegraphics[scale=0.6]{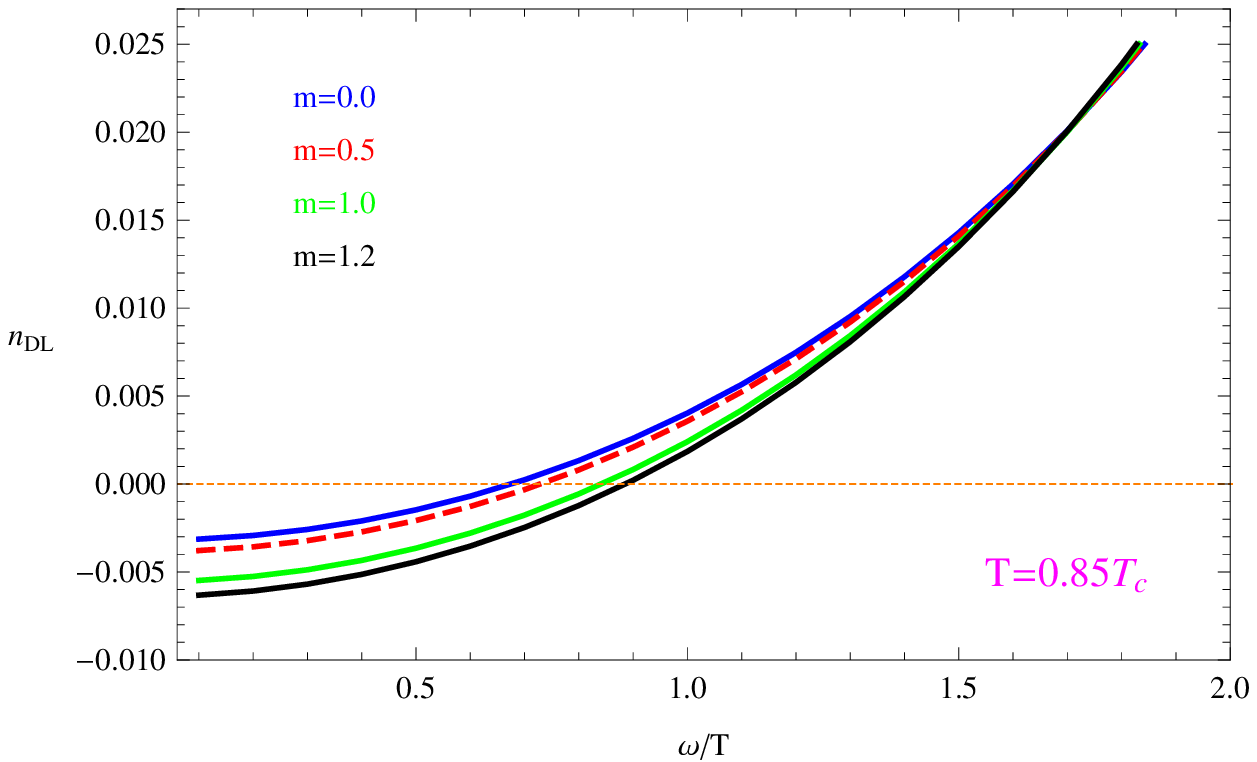}\\ \vspace{0.0cm}
\caption{\label{gravitonmassDLindex} (Color online) Depine-Lakhtakia
index $n_{DL}$ as a function of $\omega/T$ with the fixed
temperatures $T=0.81T_{c}$ (left) and $T=0.85T_{c}$ (right) for
different values of the graviton mass parameter, i.e., $m=0.0$
(blue), $0.5$ (red dotted), $1.0$ (green) and $1.2$ (black),
respectively. We have set the coupling parameter $\eta=0$ in the
numerical computation. }
\end{figure}

In order to obtain the effect of the graviton mass $m$ on the
negative refraction, we plot the behavior of the DL index $n_{DL}$
as a function of $\omega/T$ with the fixed temperatures
$T=0.81T_{c}$ and $T=0.85T_{c}$ for different values of $m$ in Fig.
\ref{gravitonmassDLindex}. Obviously, we observe the existence of
the negative DL index below a certain value of $\omega/T$ with the
fixed temperatures $T$ for all values of $m$ considered here. For
the fixed temperature, we see that the critical value of $\omega/T$,
below which the negative $n_{DL}$ appears, increases with the
increase of the graviton mass $m$, which implies that the larger
graviton mass makes the range of frequencies larger for which
negative refraction is allowed. On the other hand, we notice that
the emergence of the negative DL index becomes more obvious for the
larger graviton mass in both cases, i.e., $T=0.81T_{c}$ and
$T=0.85T_{c}$, which indicates that the system with the larger
graviton mass has a negative DL index in the superconducting phase
even $T<0.81T_{c}$ or $T>0.85T_{c}$. This means that the larger
graviton mass also makes the range of temperatures larger for which
negative refraction appears. Thus, it is interesting to note that we
can use the graviton mass $m$ to accommodate the range of
frequencies and the range of temperatures for which negative
refraction occurs.

\begin{figure}[ht]
\includegraphics[scale=0.6]{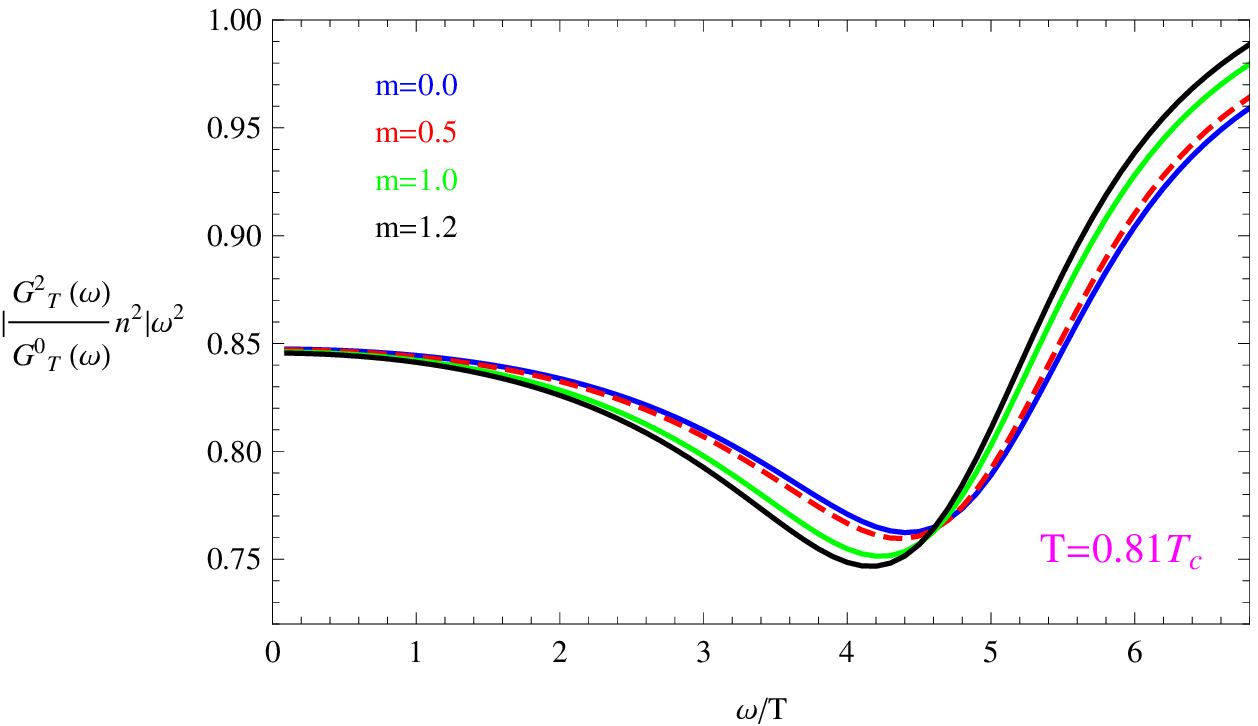}\hspace{0.2cm}%
\includegraphics[scale=0.6]{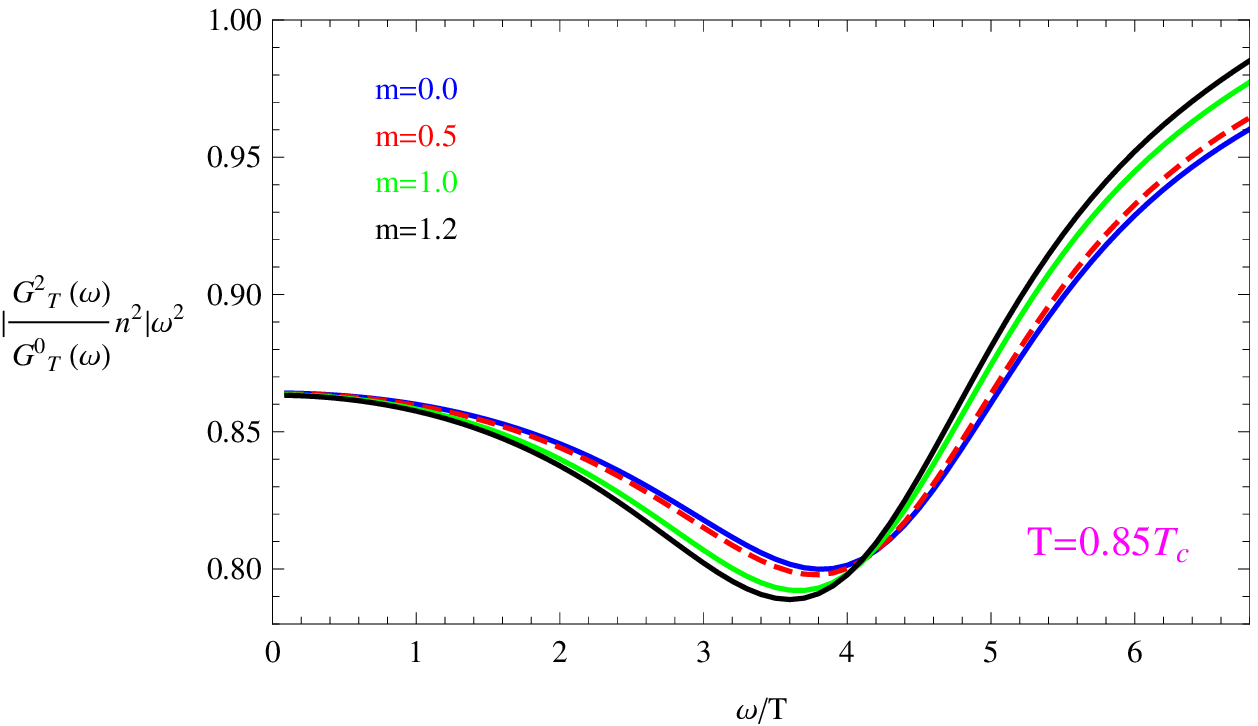}\\ \vspace{0.0cm}
\caption{\label{gravitonmassConstraint} (Color online)
$|\frac{G^{2}_{T}(\omega)}{G^{0}_{T}(\omega)}n^{2}|\omega^{2}$ as a
function of $\omega/T$ with the fixed temperatures $T=0.81T_{c}$
(left) and $T=0.85T_{c}$ (right) for different values of the
graviton mass parameter, i.e., $m=0.0$ (blue), $0.5$ (red dotted),
$1.0$ (green) and $1.2$ (black), respectively. We have set the
coupling parameter $\eta=0$ in the numerical computation. }
\end{figure}

Note that $|k^{2}G^{2}_{T}(\omega)/G^{0}_{T}(\omega)|\ll1$ is the
validity condition of Eq. (\ref{GTExpand}). With the help of Eq.
(\ref{RefractiveIndex}), we present in Fig.
\ref{gravitonmassConstraint} the curves of
$|\frac{G^{2}_{T}(\omega)}{G^{0}_{T}(\omega)}n^{2}|\omega^{2}$ as a
function of $\omega/T$ with the fixed temperatures $T=0.81T_{c}$ and
$T=0.85T_{c}$ for different values of the graviton mass parameter
$m$. Due to the appearance of a negative imaginary part of the
magnetic permeability, the validity condition is not very strictly
satisfied, just as shown in Fig. \ref{gravitonmassConstraint}, which
may be an unfortunate feature in the probe limit
\cite{MahapatraJHEP2014,DeyMT2014}. However, we find that within the
negative refraction frequency range
$|\frac{G^{2}_{T}(\omega)}{G^{0}_{T}(\omega)}n^{2}|\omega^{2}$ is
always less than one for the fixed temperatures $T$ and all values
of $m$ considered here, which means that the validity condition gets
marginally satisfied. Thus, our results are reliable.

\begin{figure}[ht]
\includegraphics[scale=0.6]{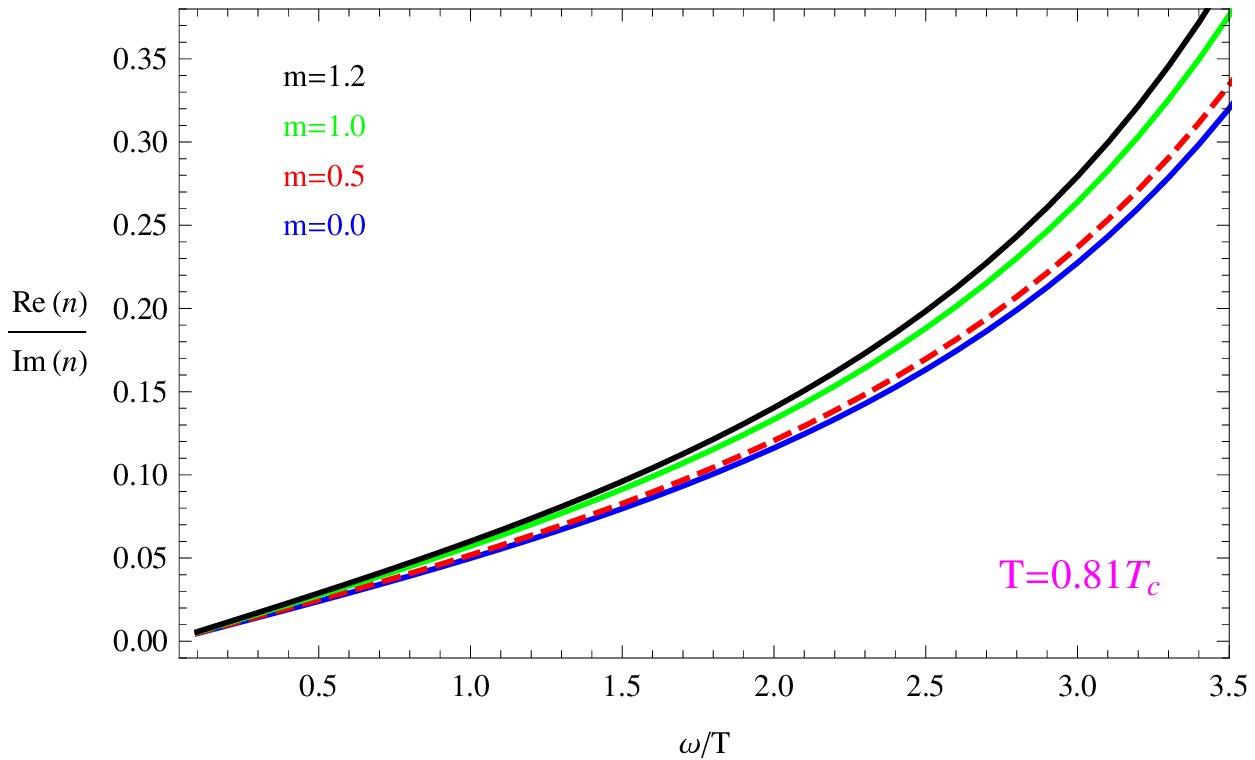}\hspace{0.2cm}%
\includegraphics[scale=0.6]{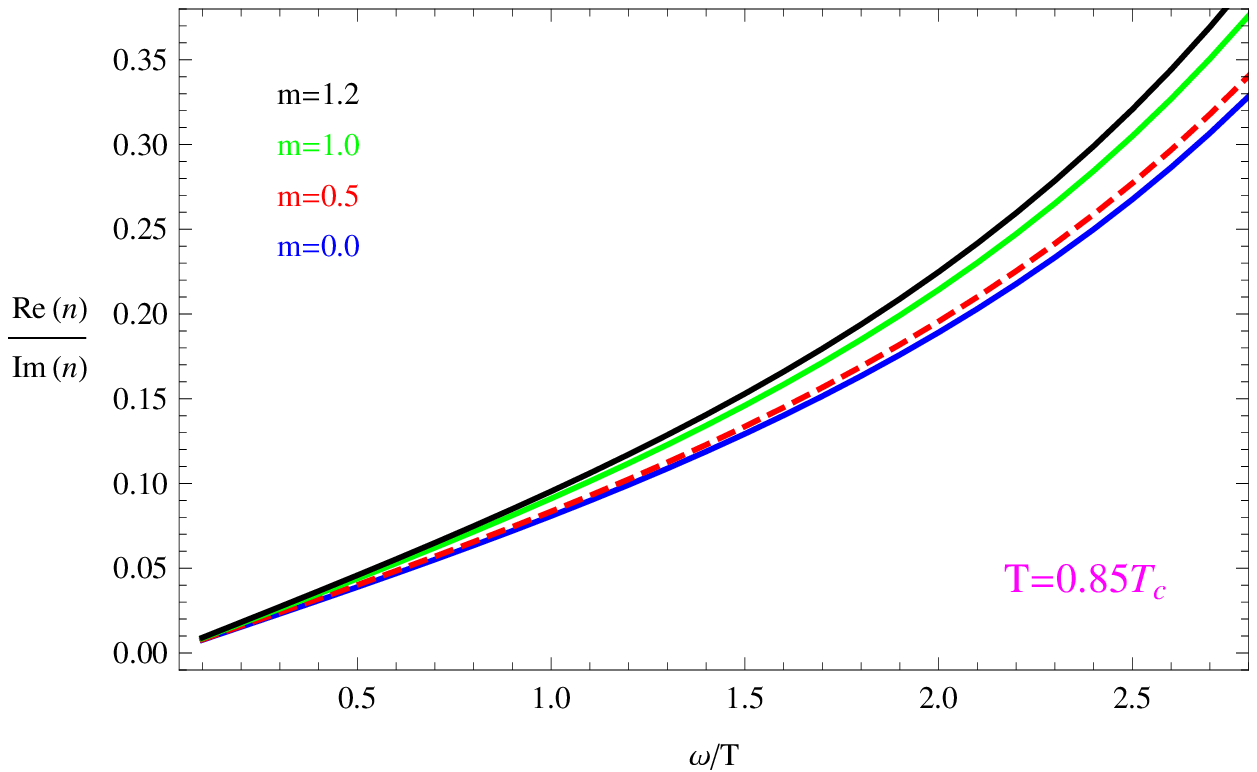}\\ \vspace{0.0cm}
\caption{\label{gravitonmassDissipation} (Color online) Ratio
Re($n$)/Im($n$) as a function of $\omega/T$ with the fixed
temperatures $T=0.81T_{c}$ (left) and $T=0.85T_{c}$ (right) for
different values of the graviton mass parameter, i.e., $m=0.0$
(blue), $0.5$ (red dotted), $1.0$ (green) and $1.2$ (black),
respectively. We have set the coupling parameter $\eta=0$ in the
numerical computation. }
\end{figure}

It is also important to calculate the ratio Re($n$)/Im($n$) and
analyze the propagation Re($n$) and the dissipation Im($n$) of
electromagnetic waves in systems. In Fig.
\ref{gravitonmassDissipation}, we plot the ratio Re($n$)/Im($n$) as
a function of $\omega/T$ with the fixed temperatures $T=0.81T_{c}$
and $T=0.85T_{c}$ for different values of the graviton mass
parameter $m$. In both cases, i.e., $T=0.81T_{c}$ and $T=0.85T_{c}$,
we observe that the magnitude of Re($n$)/Im($n$) is small in the
frequency range where $n_{DL}$ is negative, which indicates large
dissipation in the system. However, for the fixed temperature, the
ratio increases with increasing values of $m$ for the fixed
$\omega/T$, which means that the larger graviton mass does enhance
the magnitude of Re($n$)/Im($n$) and thereby increases the
propagation. Hence we can use the graviton mass to reduce the
dissipation and improve the propagation in the holographic setup.

\subsection{Effect of the Einstein tensor on the negative refractive index}

\begin{figure}[ht]
\includegraphics[scale=0.6]{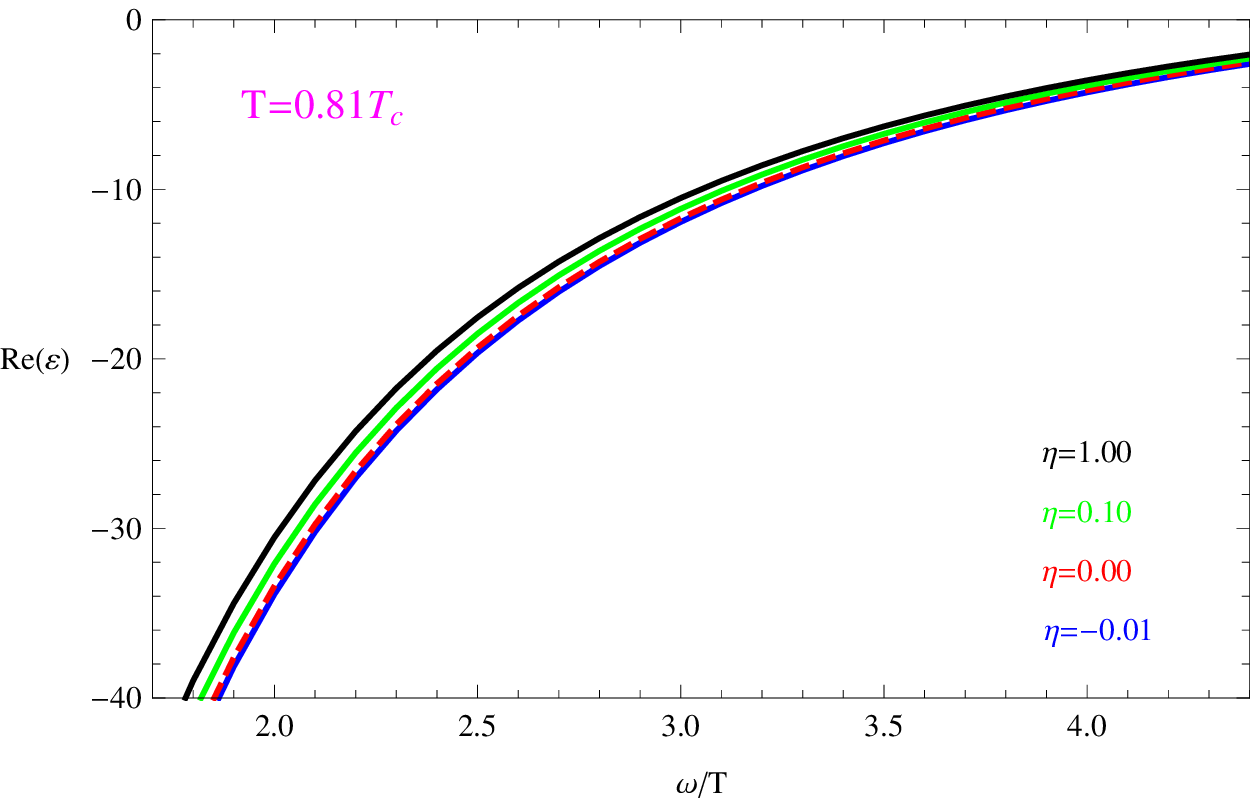}\hspace{0.2cm}%
\includegraphics[scale=0.6]{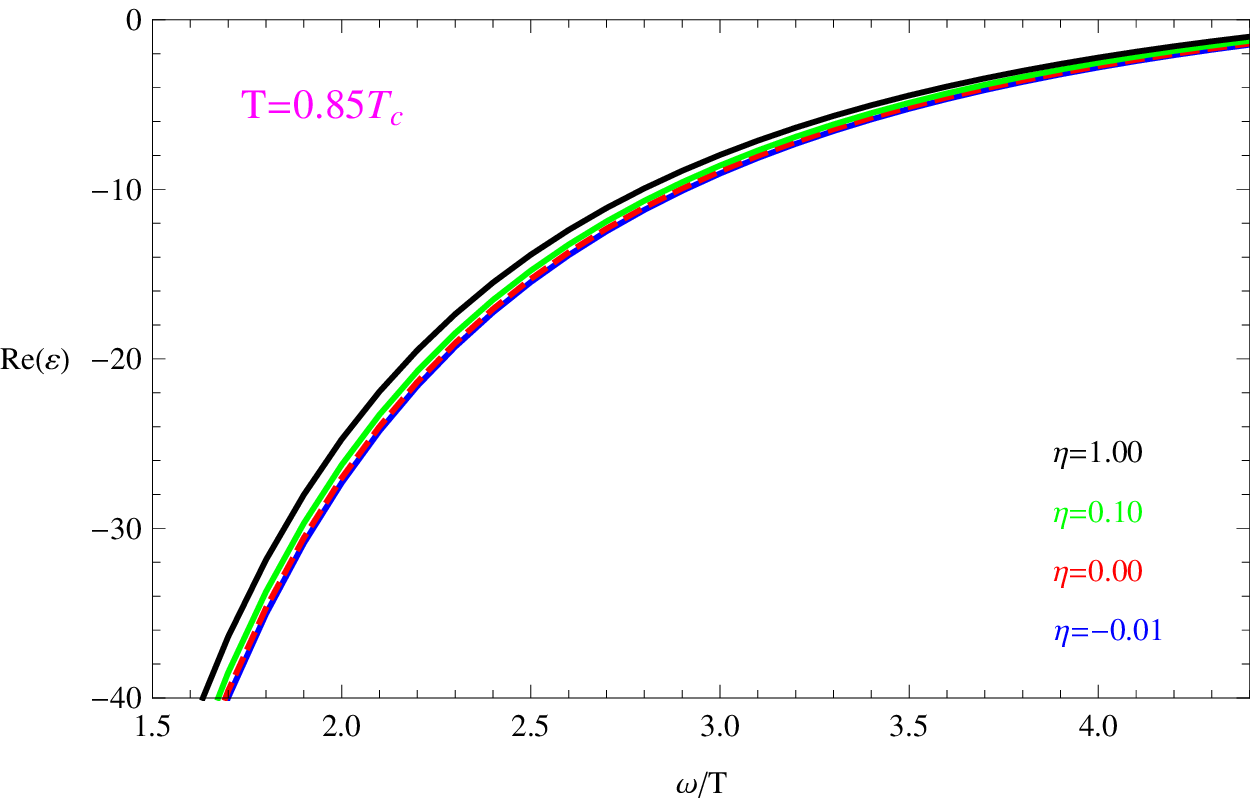}\\ \vspace{0.0cm}
\includegraphics[scale=0.6]{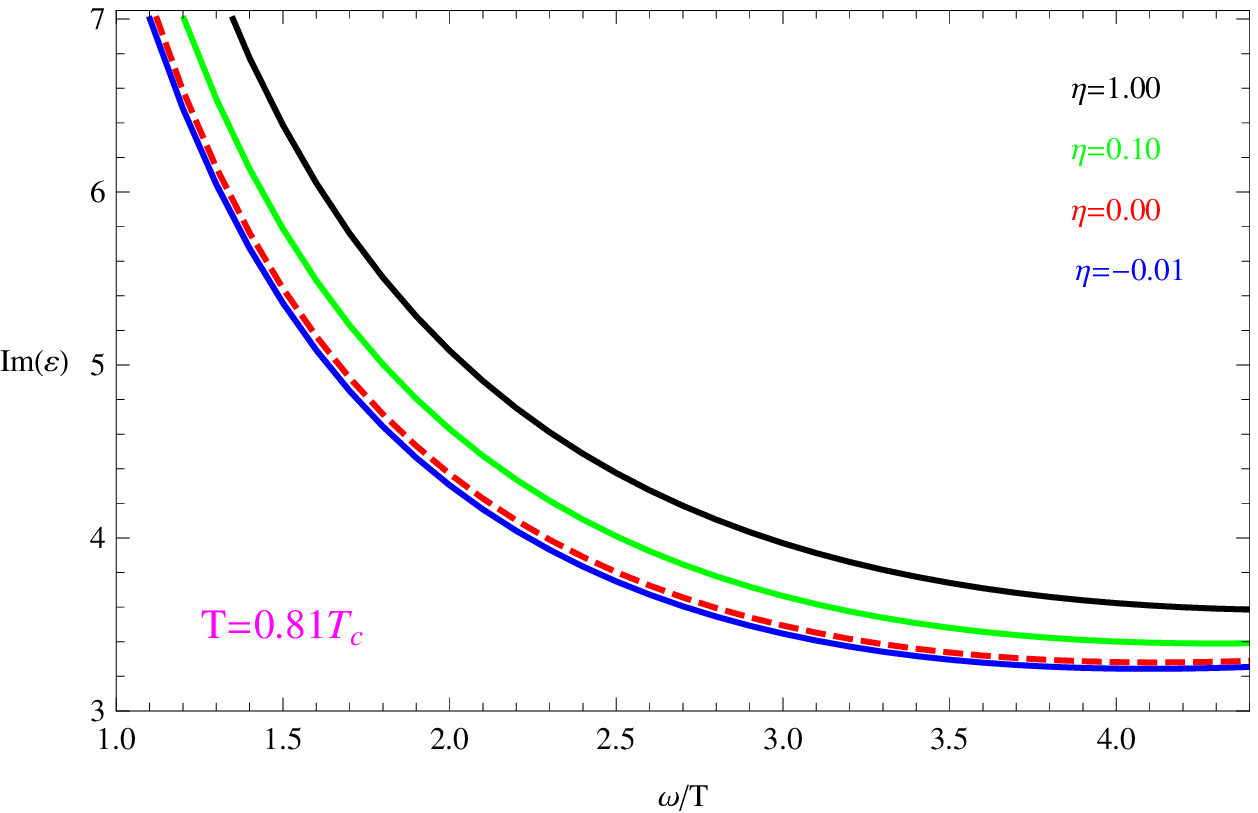}\hspace{0.2cm}%
\includegraphics[scale=0.6]{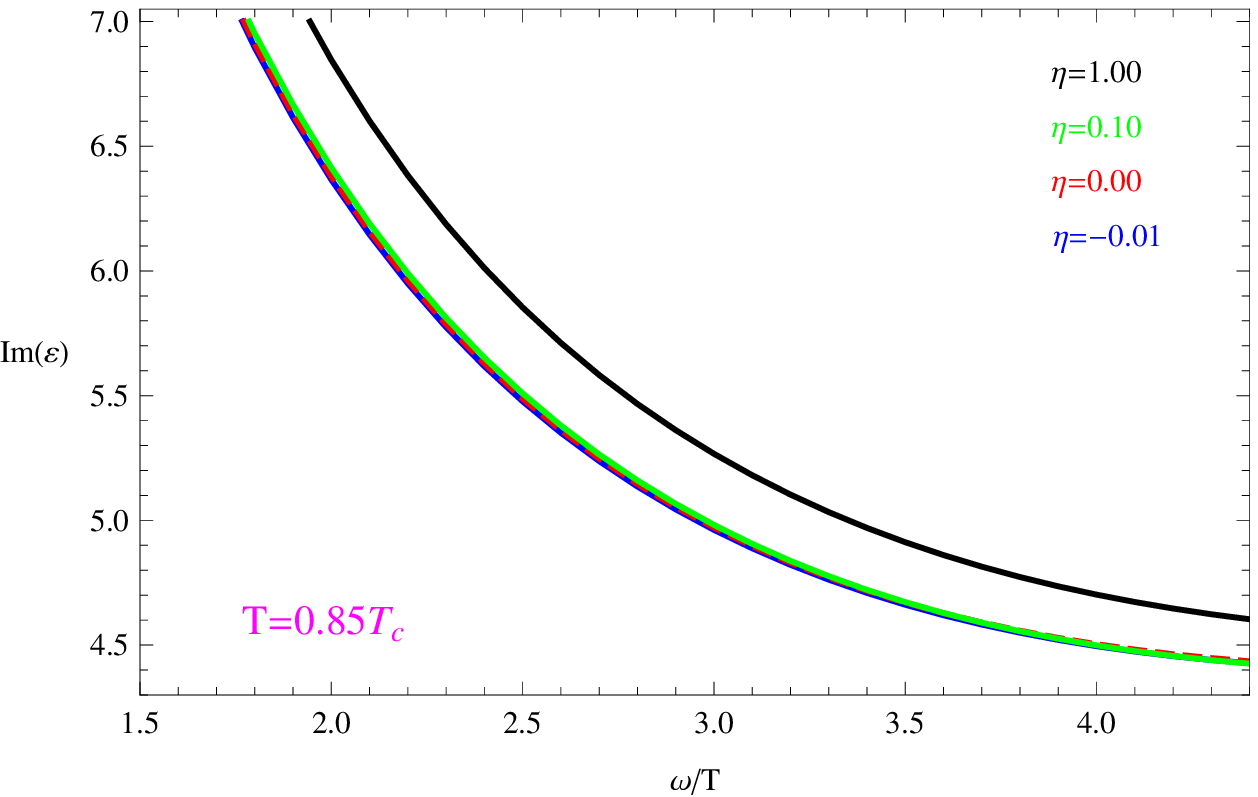}\\ \vspace{0.0cm}
\caption{\label{EinsteintensorPermittivity} (Color online) Real
(top) and imaginary (bottom) parts of the permittivity $\epsilon$ as
a function of $\omega/T$ with the fixed temperatures $T=0.81T_{c}$
(left) and $T=0.85T_{c}$ (right) for different values of the
coupling parameter, i.e., $\eta=-0.01$ (blue), $0.00$ (red dotted),
$0.01$ (green) and $1.00$ (black), respectively. We have set the
graviton mass $m=1.0$ in the numerical computation. }
\end{figure}

\begin{figure}[ht]
\includegraphics[scale=0.6]{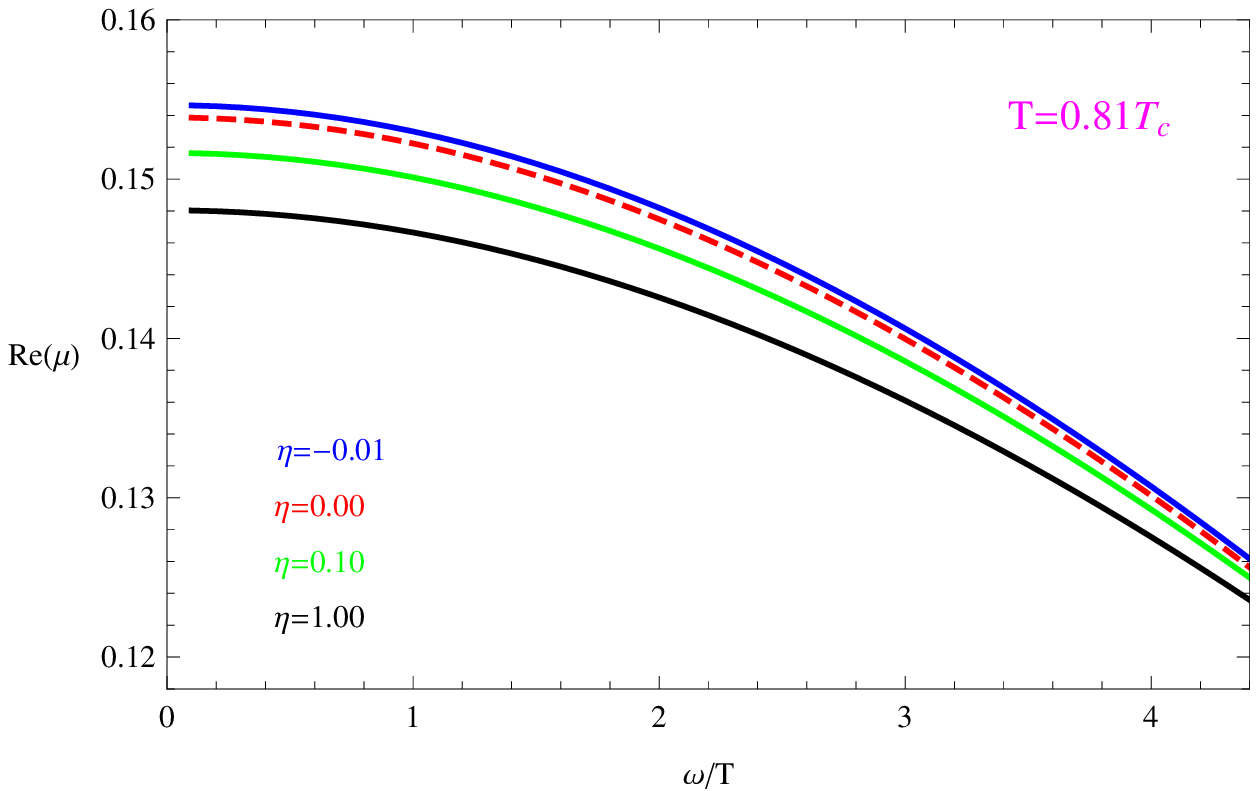}\hspace{0.2cm}%
\includegraphics[scale=0.6]{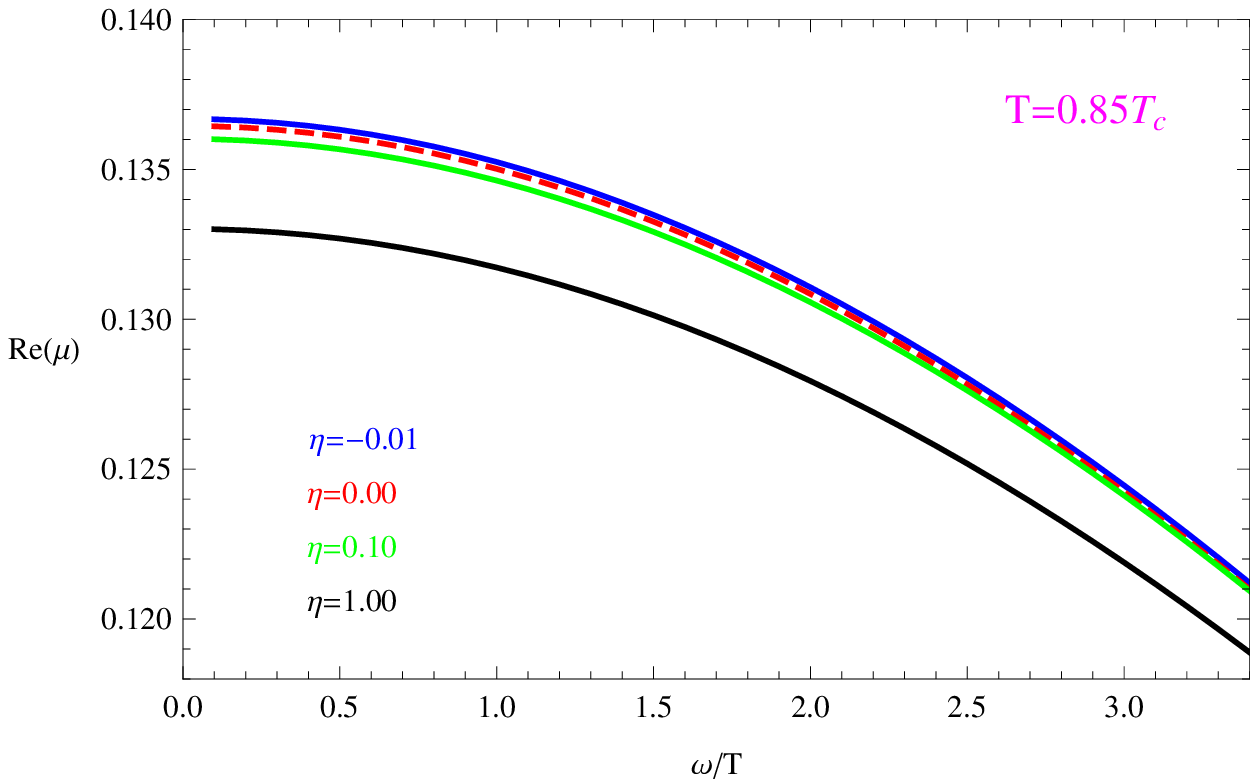}\\ \vspace{0.0cm}
\includegraphics[scale=0.6]{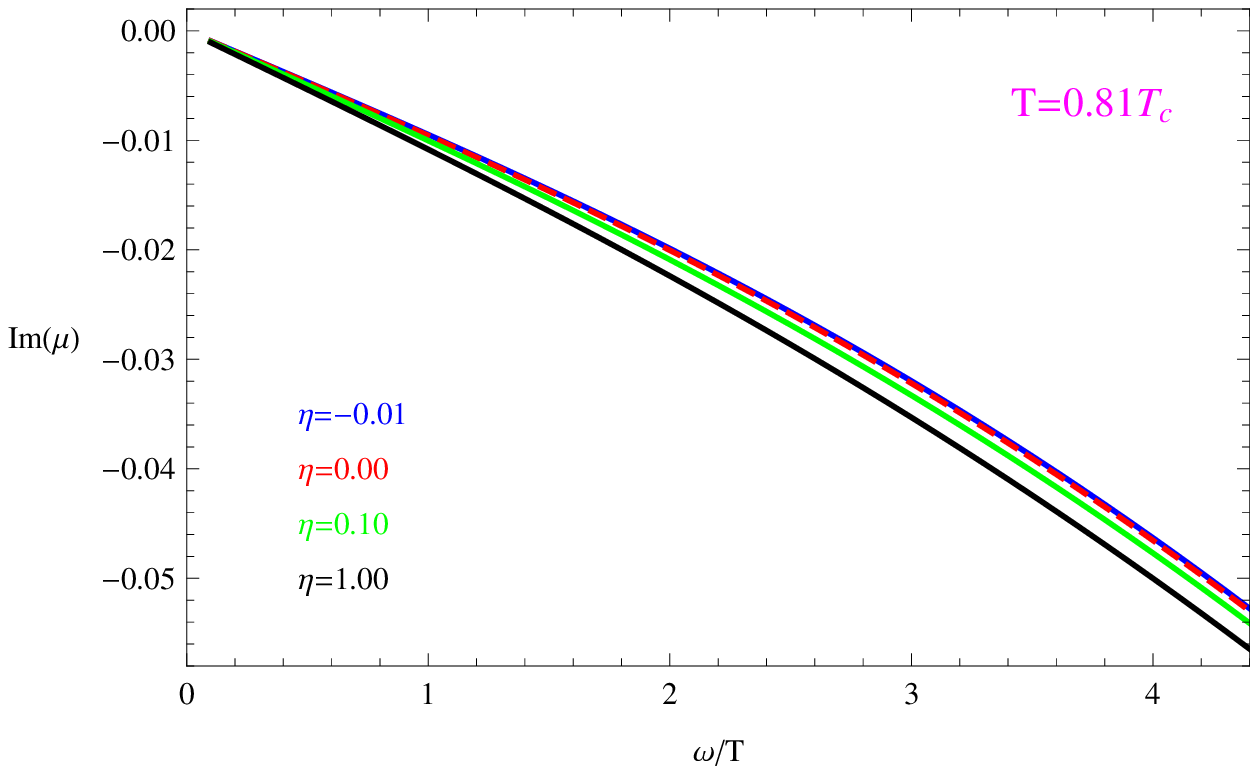}\hspace{0.2cm}%
\includegraphics[scale=0.6]{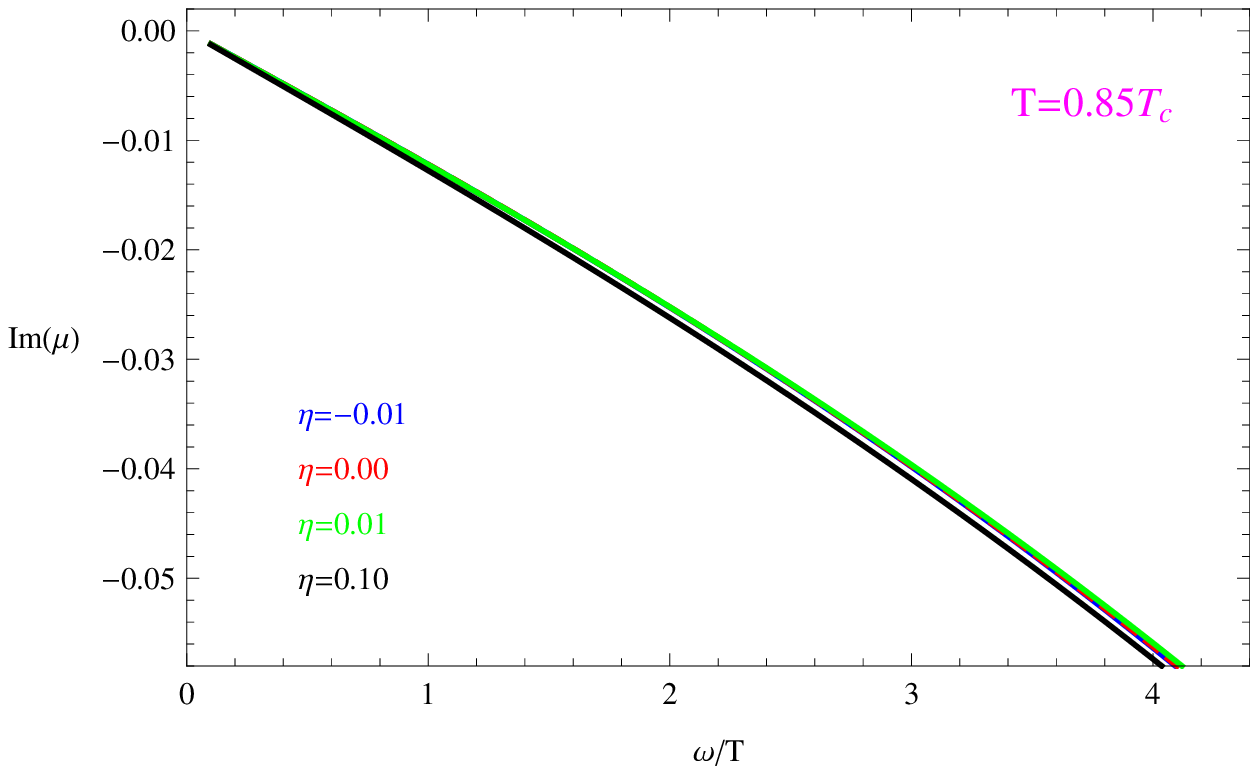}\\ \vspace{0.0cm}
\caption{\label{EinsteintensorPermeability} (Color online) Real
(top) and imaginary (bottom) parts of the permeability $\mu$ as a
function of $\omega/T$ with the fixed temperatures $T=0.81T_{c}$
(left) and $T=0.85T_{c}$ (right) for different values of the
coupling parameter, i.e., $\eta=-0.01$ (blue), $0.00$ (red dotted),
$0.01$ (green) and $1.00$ (black), respectively. We have set the
graviton mass $m=1.0$ in the numerical computation. }
\end{figure}

Moving on to the effect of the Einstein tensor on the negative
refraction, we firstly give the behaviors of the permittivity
$\epsilon$ and permeability $\mu$ as a function of $\omega/T$ with
the fixed temperatures $T=0.81T_{c}$ and $T=0.85T_{c}$ for different
values of the Einstein tensor coupling parameter in case of the
graviton mass $m=1.0$ in Figs. \ref{EinsteintensorPermittivity} and
\ref{EinsteintensorPermeability}. It is shown that, for the
permittivity, Re$(\epsilon)$ is negative at low frequencies and
Im$(\epsilon)$ is always positive, but for the permeability,
Re$(\mu)$ is always positive and Im$(\mu)$ is negative, which
implies that the refraction may be negative since the real parts of
the permittivity Re$(\epsilon)$ and of the permeability Re$(\mu)$
are not simultaneously negative. It should be noted that we have
defined an effective magnetic permeability that is not an
observable, although there is a caveat, i.e., the appearance of a
negative imaginary part of the magnetic permeability. We will
proceed with this point in mind.

\begin{figure}[ht]
\includegraphics[scale=0.6]{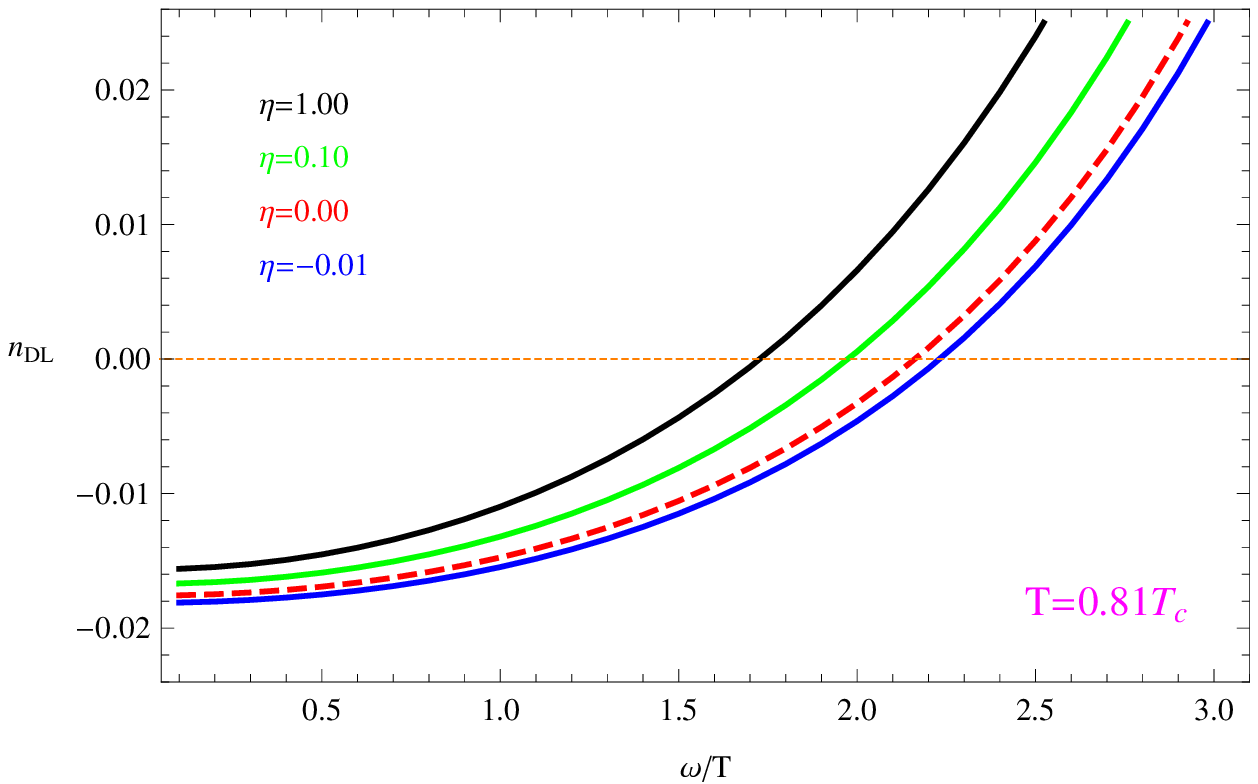}\hspace{0.2cm}%
\includegraphics[scale=0.6]{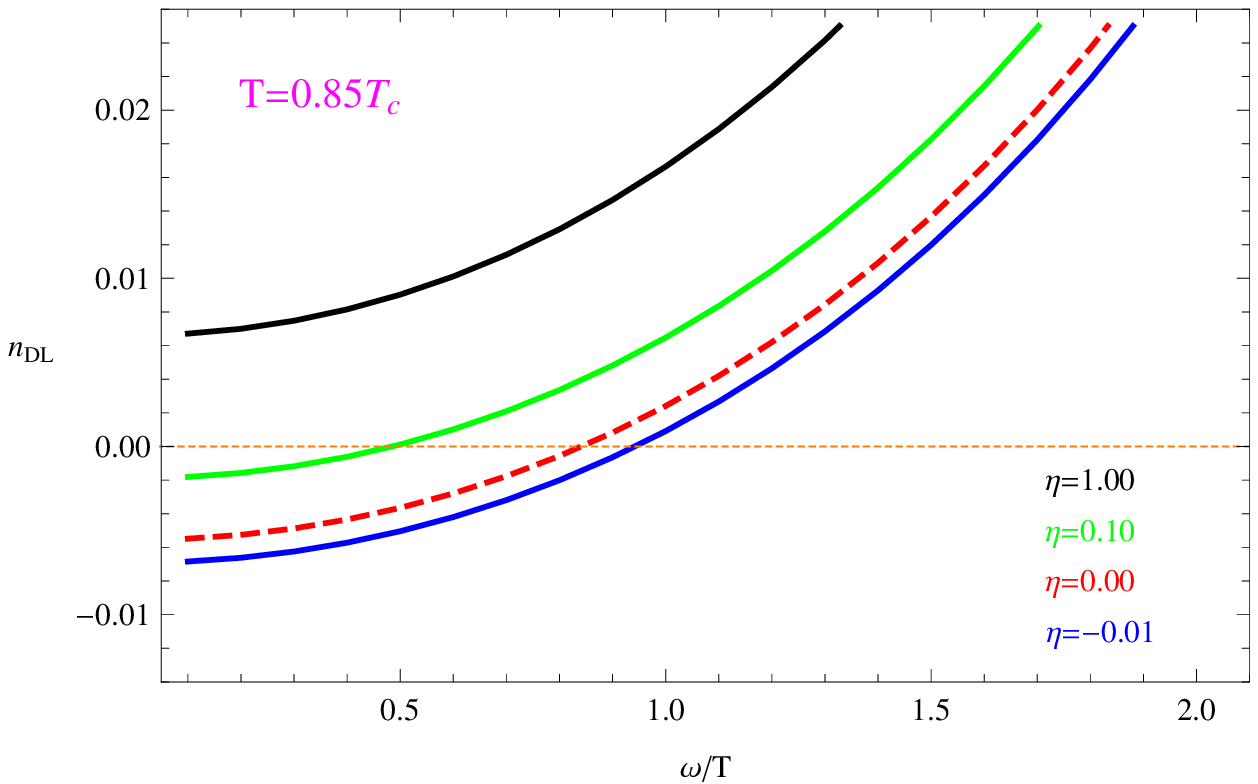}\\ \vspace{0.0cm}
\caption{\label{EinsteintensorDLindex} (Color online)
Depine-Lakhtakia index $n_{DL}$ as a function of $\omega/T$ with the
fixed temperatures $T=0.81T_{c}$ (left) and $T=0.85T_{c}$ (right)
for different values of the coupling parameter, i.e., $\eta=-0.01$
(blue), $0.00$ (red dotted), $0.01$ (green) and $1.00$ (black),
respectively. We have set the graviton mass $m=1.0$ in the numerical
computation. }
\end{figure}

In Fig. \ref{EinsteintensorDLindex}, we present the curves of the DL
index $n_{DL}$ as a function of $\omega/T$ with the fixed
temperatures $T=0.81T_{c}$ and $T=0.85T_{c}$ for different values of
the coupling parameter $\eta$ in case of the graviton mass $m=1.0$.
For the fixed temperature, we observe that the critical value of
$\omega/T$, below which the system has negative $n_{DL}$, decreases
with the increase of the coupling parameter $\eta$, which indicates
that the larger Einstein tensor coupling parameter reduces the range
of frequencies for which negative refraction occurs. On the other
hand, we see that the negative $n_{DL}$ appears in the case of the
coupling parameter $\eta=1.00$ at the temperature $T=0.81T_{c}$ and
disappears at $T=0.85T_{c}$, but the negative $n_{DL}$ always
appears in the case of the coupling parameter $\eta=0.10$
($\eta=0.00$ or $\eta=-0.01$) at the temperatures $T=0.81T_{c}$ and
$T=0.85T_{c}$, which means that the larger Einstein tensor coupling
parameter also reduces the range of temperatures for which negative
refraction occurs. The behavior of the Einstein tensor is
reminiscent of that seen for the graviton mass, so we conclude that
the graviton mass and Einstein tensor have completely different
effects on the negative refractive index of our holographic systems.

\begin{figure}[ht]
\includegraphics[scale=0.6]{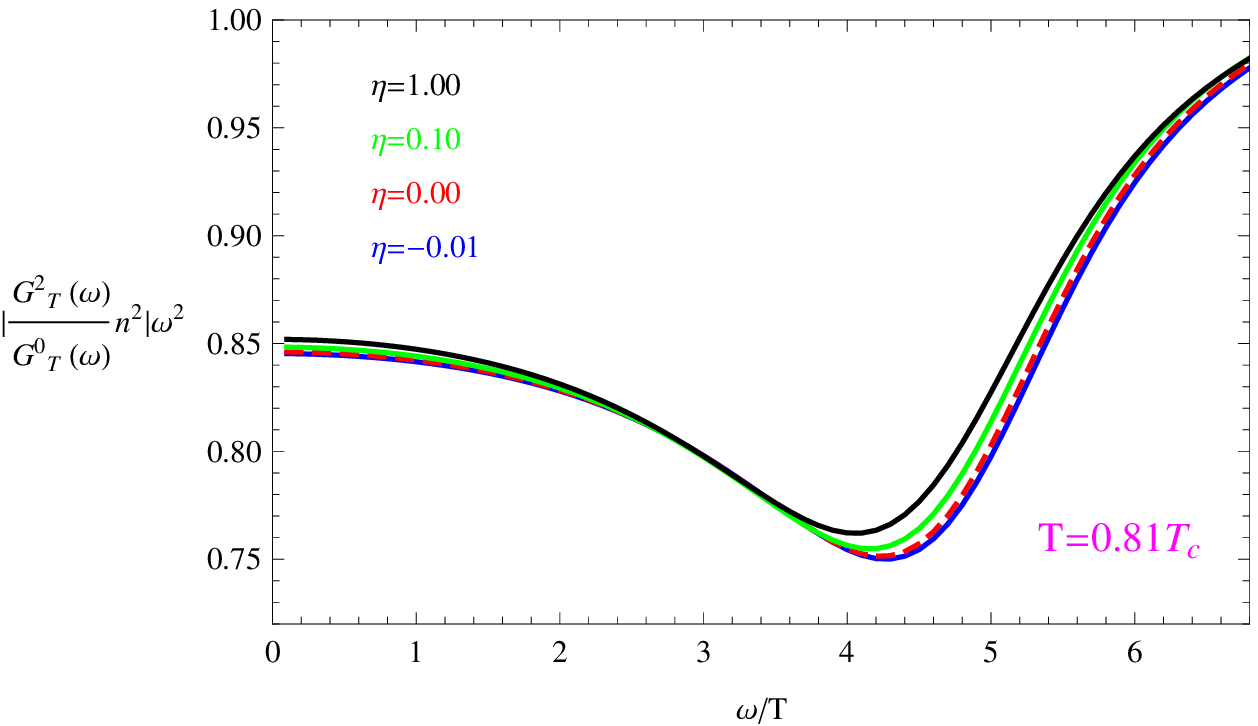}\hspace{0.2cm}%
\includegraphics[scale=0.6]{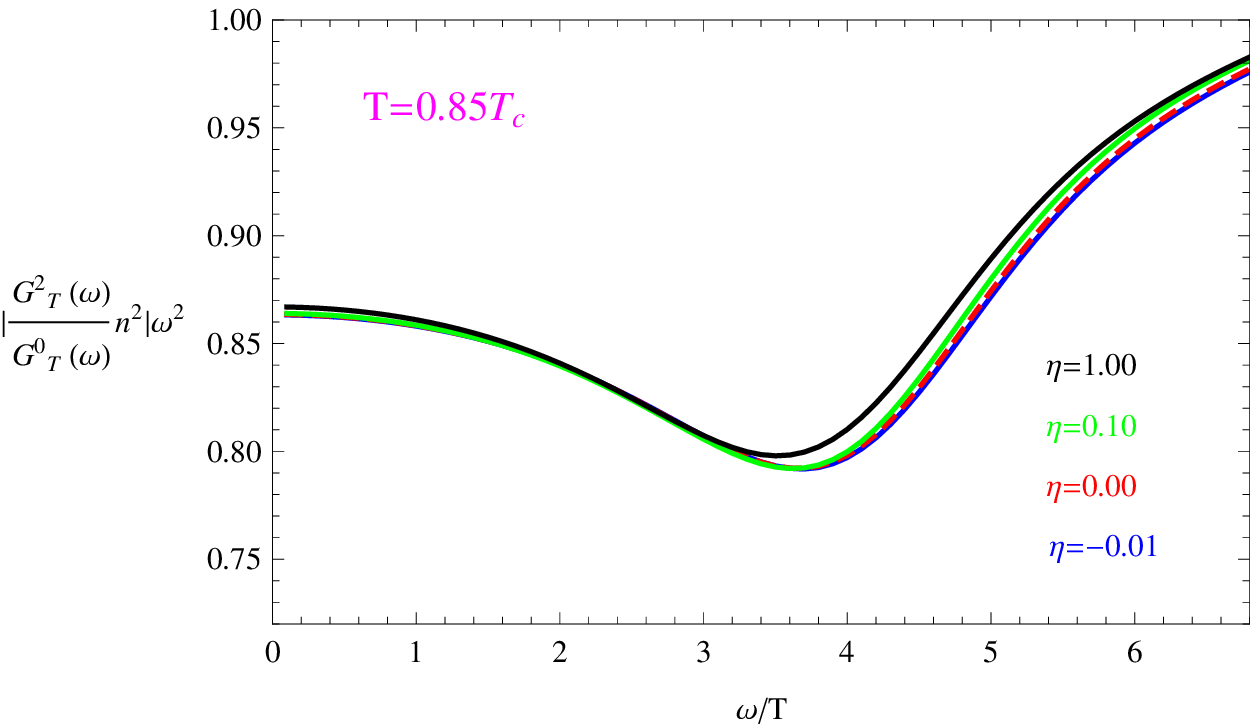}\\ \vspace{0.0cm}
\caption{\label{EinsteintensorConstraint} (Color online)
$|\frac{G^{2}_{T}(\omega)}{G^{0}_{T}(\omega)}n^{2}|\omega^{2}$ as a
function of $\omega/T$ with the fixed temperatures $T=0.81T_{c}$
(left) and $T=0.85T_{c}$ (right) for different values of the
coupling parameter, i.e., $\eta=-0.01$ (blue), $0.00$ (red dotted),
$0.01$ (green) and $1.00$ (black), respectively. We have set the
graviton mass $m=1.0$ in the numerical computation. }
\end{figure}

In order to verify the validity of the expansion used in Eq.
(\ref{GTExpand}), we plot in Fig. \ref{EinsteintensorConstraint} the
behavior of
$|\frac{G^{2}_{T}(\omega)}{G^{0}_{T}(\omega)}n^{2}|\omega^{2}$ as a
function of $\omega/T$ with the fixed temperatures $T=0.81T_{c}$ and
$T=0.85T_{c}$ for different values of the coupling parameter $\eta$
in case of the graviton mass $m=1.0$. It is shown that, within the
negative refraction frequency range, the constraint
$|k^{2}G^{2}_{T}(\omega)/G^{0}_{T}(\omega)|\ll1$ is marginally
satisfied for all values of $\eta$ considered here since the
$\epsilon-\mu$ analysis is valid only for the frequencies for which
this constraint is not violated.

\begin{figure}[ht]
\includegraphics[scale=0.6]{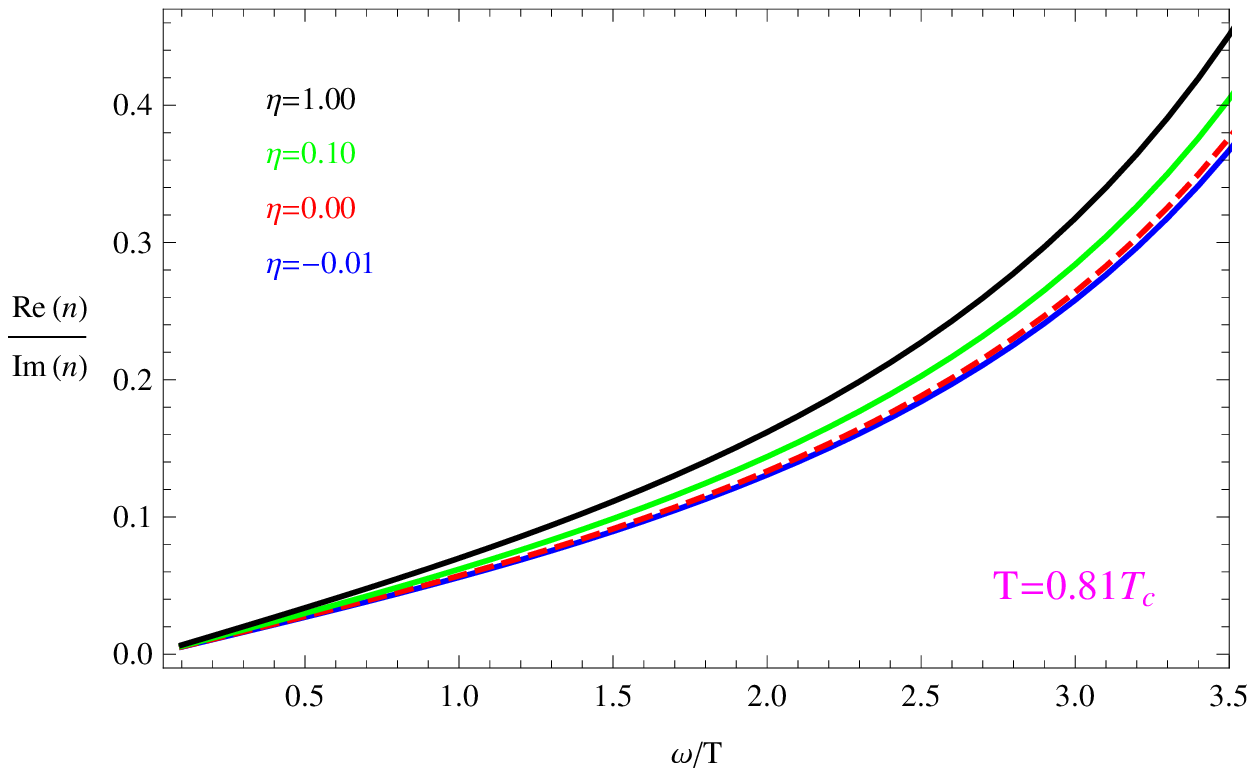}\hspace{0.2cm}%
\includegraphics[scale=0.6]{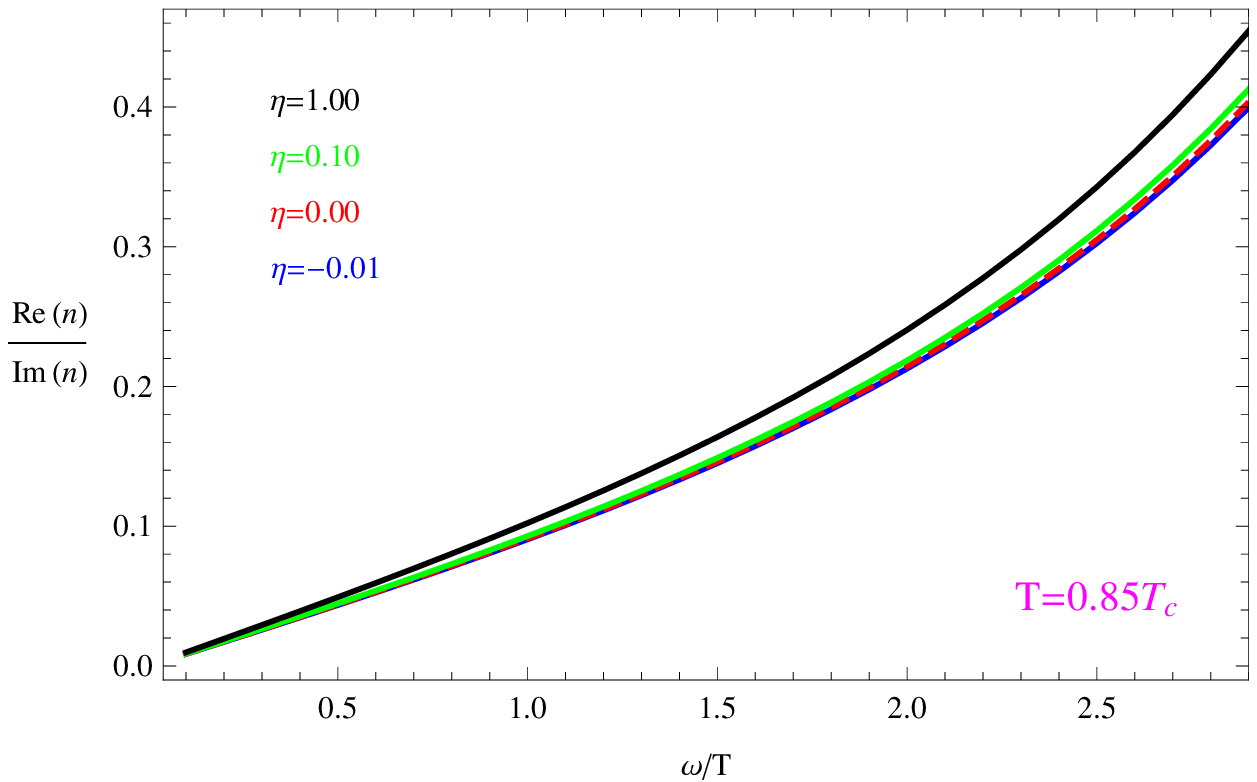}\\ \vspace{0.0cm}
\caption{\label{EinsteintensorDissipation} (Color online) Ratio
$Re(n)/Im(n)$ as a function of $\omega/T$ with the fixed
temperatures $T=0.81T_{c}$ (left) and $T=0.85T_{c}$ (right) for
different values of the coupling parameter, i.e., $\eta=-0.01$
(blue), $0.00$ (red dotted), $0.01$ (green) and $1.00$ (black),
respectively. We have set the graviton mass $m=1.0$ in the numerical
computation. }
\end{figure}

Finally, we present in Fig. \ref{EinsteintensorDissipation} the
ratio Re($n$)/Im($n$) as a function of $\omega/T$ with the fixed
temperatures $T=0.81T_{c}$ and $T=0.85T_{c}$ for different values of
the coupling parameter $\eta$ in case of the graviton mass $m=1.0$.
Obviously, for the fixed temperature, i.e., $T=0.81T_{c}$ or
$T=0.85T_{c}$, the ratio increases with increasing values of $\eta$
for the fixed $\omega/T$, which indicates that the larger Einstein
tensor coupling parameter enhances the magnitude of Re($n$)/Im($n$)
and can be used to improve the propagation in the holographic setup.
Interestingly, the effect of the Einstein tensor on the ratio
Re($n$)/Im($n$) is similar to that of the graviton mass.

\section{conclusions}

Recent studies showed that the graviton mass parameter has the
similar effect on the phase transition from superconductivity to a
normal metal as that of the doping \cite{HuPRD2016} and the coupling
of a scalar field to the Einstein tensor in the gravity bulk can
have a dual interpretation on the boundary corresponding to
impurities concentrations in a real material \cite{KuangE2016}. In
order to investigate systematically the influence of the doping or
impurity on the superconductors in strongly coupled condensed matter
systems by holography, we constructed the generalized
superconductors from the coupling of a scalar field to the Einstein
tensor in the massive gravity and analyzed the effects of the
graviton mass and Einstein tensor on the holographic superconductor
models and their negative refraction. In the probe limit, we found
that the critical chemical potential increases with the increase of
the graviton mass or Einstein tensor coupling parameter, which
supports the findings in Ref. \cite{HuPRD2016,KuangE2016} and
indicates that the larger graviton mass or Einstein tensor
parameters hinder the formation of the condensation. Interestingly,
in our model we observed that the second-order transition occurs
before the first-order transition to a new superconducting phase
when the chemical potential increases (namely the temperature
decreases), i.e, the so-called Cave of Winds, and the larger
graviton mass but smaller coupling parameter make it easier for the
emergence of the Cave of Winds. On the other hand, we investigated
the optical properties of our generalized holographic
superconductors and found the existence of the negative
Depine-Lakhtakia index in the superconducting phase at small
frequencies. We noted that the larger graviton mass but smaller
Einstein tensor coupling parameter make the range of frequencies or
the range of temperatures larger for which a negative DL index
occurs, which means that the graviton mass and Einstein tensor have
completely different effects on the negative refraction. Considering
the diverse effects of the doping or impurity on the refractive
index of the real material, for example LiNbO$_{3}$
\cite{Minakata,PHKAK}, our result suggests that the doping or
impurity will have various effects on the negative refraction in the
high-temperature superconductor systems. Furthermore, we found that
the larger graviton mass and coupling parameters both can reduce the
dissipation and improve the propagation in the holographic setup.
Our analysis shows that the graviton mass and the coupling of the
scalar field to the Einstein tensor both can play important roles in
determining the optical properties of the boundary theory.

\begin{acknowledgments}

This work was supported by the National Natural Science Foundation
of China under Grant Nos. 11775076, 11475061 and 11690034; Hunan
Provincial Natural Science Foundation of China under Grant No.
2016JJ1012.

\end{acknowledgments}

\end{document}